\documentclass[aps,prb,preprint,superscriptaddress,showpacs,showkeys,11pt,a4paper]{revtex4} 
\usepackage{amsfonts}
\usepackage{amssymb}
\usepackage{amsmath}
\usepackage{lipsum}
\usepackage{graphicx}
\usepackage{fixfoot}
\usepackage{color}
\usepackage{rotating}
\begin{document} 

%\title{Magnetoimpedance effect for the thin film geometry in a wide frequency range: Numerical calculation and experiment}
\title{Magnetoimpedance effect at the high frequency range for the thin film geometry: Numerical calculation and experiment}

\author{M.~A.~Corr\^{e}a} 
\email[Electronic address: ]{marciocorrea@dfte.ufrn.br}
\affiliation{Departamento de F\'{i}sica Te\'{o}rica e Experimental, Universidade Federal do Rio Grande do Norte, 59078-900 Natal, RN, Brazil} 
\author{F. Bohn} 
\affiliation{Departamento de F\'{i}sica Te\'{o}rica e Experimental, Universidade Federal do Rio Grande do Norte, 59078-900 Natal, RN, Brazil} 
\author{R. B. da Silva} 
\affiliation{Departamento de F\'{i}sica, Universidade Federal de Santa Maria, 97105-900 Santa Maria, RS, Brazil} 
\author{R. L. Sommer} 
\affiliation{Centro Brasileiro de Pesquisas F\'{i}sicas, Rua Dr.\ Xavier Sigaud 150, Urca, 22290-180 Rio de Janeiro, RJ, Brazil} 

\date{\today} 

\begin{abstract} 
The magnetoimpedance effect is a versatile tool to investigate ferromagnetic materials, revealing aspects on the fundamental physics associated to magnetization dynamics, broadband magnetic properties, important issues for current and emerging technological applications for magnetic sensors, as well as insights on ferromagnetic resonance effect at non-saturated magnetic states. Here, we perform a theoretical and experimental investigation of the magnetoimpedance effect for the thin film geometry in a wide frequency range. We calculate the longitudinal magnetoimpedance for single layered, multilayered or exchange biased systems from an approach that considers a magnetic permeability model for planar geometry and the appropriate magnetic free energy density for each structure. From numerical calculations and experimental results found in literature, we analyze the magnetoimpedance behavior, and discuss the main features and advantages of each structure. To test the robustness of the approach, we directly compare theoretical results with experimental magnetoimpedance measurements obtained in a wide range of frequencies for an exchange biased multilayered film. Thus, we provide experimental evidence to confirm the validity of the theoretical approach employed to describe the magnetoimpedance in ferromagnetic films, revealed by the good agreement between numerical calculations and experimental results.
\end{abstract} 

\pacs{75.40.Gb, 75.30.Gw, 75.60.-d} 

\keywords{tensor magnetoimpedance, thin film, dynamics magnetization} 

\maketitle 

\section{Introduction} 
\label{Introduction}

The study of dynamical phenomena has provided central advances on magnetization dynamics during the past decades. Usually, the investigations are based on traditional ferromagnetic resonance (FMR) experiments, in which the sample is submitted to an intense external magnetic field, saturating it magnetically. From FMR measurements, information regarding magnetic anisotropies, damping parameter and other important parameters related to the magnetic dynamics can be reached. However, nowadays, similar information can be accessibly obtained also through the study of the magnetoimpedance effect. This effect is a versatile tool commonly employed to investigate ferromagnetic materials, revealing aspects on the fundamental physics associated to magnetization dynamics, broadband magnetic properties~\cite{PRB81p134421, JAP115p103908, APEX6p093001}, as well as on important issues for current and emerging technological applications for magnetic sensors~\cite{SAAP204p20, IEEETM49p3874, JAP98p014316, JMMM293p671,SAAP106p187}. Besides, further insights on FMR effect at non-saturated magnetic states of the sample can be easily gotten, making possible the study of local resonances and their influence in the dynamics magnetization.

The magnetoimpedance effect (MI) corresponds to the change of the real and imaginary components of electrical impedance of a ferromagnetic sample caused by the action of an external static magnetic field. In a typical MI experiment, the studied sample is also submitted to an alternate magnetic field associated to the electric current $I_{ac} = I_o \exp (i2 \pi ft)$, $f$ being the probe current frequency. Irrespective to the sample geometry, the overall effect of these magnetic fields is to induce strong modifications of the effective magnetic permeability. 

Experimentally, studies on MI have been widely performed in sheets~\cite{JAP84p3792}, magnetic ribbons~\cite{PRB51p3926, JAP79p5462, PRB53pR5982, JAP79p6546, PRB60p6685, JAP95p1364}, wires~\cite{APL64p3652, APL65p1189, PRB50p16737, IEEETM31p1266, APL77p121, SAA59p20, JAP91p7436, JPDAP40p3233}, and in ferromagnetic films with several structures, such as single layered~\cite{APL67p3346, JAP101p033908,JAP115p17a303}, multilayered~\cite{SAA129p256, APL94p042501, APL96p232501, JPDAP43p295004, TSF520p2173, JMMM355p136, JAP115p103908, APL105p102409}, and structured multilayered samples~\cite{IEEETM32p4965, JMMM242p291, JPCM16p6561, JPDAP41p175003, JPDAP43p295004, JAP101p043905, APL94p042501, APL96p232501, APL104p102405, APL105p102409}.

The general theoretical approach to the MI problem focuses on its determination as a function of magnetic field for a range of frequencies. Traditionally, the changes of magnetic permeability and impedance with magnetic fields at different frequency ranges are caused by three distinct mechanisms~\cite{PMS53p323,IEEETM29p1245,APL64p3652,APL69p3084}: magnetoinductive effect, skin effect, and FMR effect. Thus, MI can generally be classified into the three frequency regimes~\cite{JAP110p093914}.
%The general theoretical approach to the MI problem focuses on its determination as a function of magnetic field for a range of frequencies. Thus, MI can generally be classified into the three frequency regimes~\cite{PMS53p323}. For the low-frequency range, MI schematization was proposed by Mohri et al.~\cite{IEEETM29p1245}, considering changes in the inductive part of impedance and this is known as the magnetoinductive effect. At the intermediate frequency range, Beach et al.~\cite{APL64p3652} considered that the skin effect was responsible for transverse magnetic permeability changes and consequently, for variations in the magnetoimpedance effect. Finally, at high frequencies, Yelon et al.ii demonstrated that field configurations favor the appearance of ferromagnetic resonance, the main mechanism responsible for magnetic permeability changes and MI variations. ~\cite{APL69p3084}
Moreover, the MI behavior with magnetic field and probe current frequency becomes more complex, since it also depends on magnetic properties, such as magnetic anisotropies, as well as sample dimensions and geometry. Given that distinct effects affect the magnetic permeability behavior at different frequency ranges and different properties influences the MI, the description of the magnetoimpedance effect over a wide range of frequency becomes a difficult task. For this reason, the comprehension on the theoretical and experimental point of views of the magnetoimpedance effect is fundamental for the development of new materials with optimized response.

Since the system geometry has an important role on MI results, several studies have been performed to obtain further information on this dependence. Considerable attention have been given to describe the MI effect in samples presenting cylindrical geometry with distinct anisotropy configurations~\cite{JPDAP37p2773, PRB63p144424, JAP88p379, JAP113p243902}. For this case, e. g., Makhnovskiy {\it et al.}~\cite{PRB63p144424} have reported a very strict study on the surface impedance tensor, in which theoretical results for the cylindrical geometry are directly compared to experimental measurements acquired for ferromagnetic wires. In addition, Usov {\it et al.}~\cite{JAP113p243902} have presented theoretical and experimental results for ferromagnetic wires with weak helical anisotropy. 
%{\color{blue} In both, the MI effect is explored in a limited low frequency range and do not present the high frequency effects in the MI measurements.}

Regarding the MI effect for the case of planar systems, an important study has been performed in single layers by Kraus~\cite{JMMM195p764}, who performed the calculation of the MI effect in a single planar conductor and studied the influence of the Gilbert damping constant, the angle between the anisotropy direction and the applied magnetic field on the MI effect. Moreover, Panina {\it et al.}~\cite{JAP89p7221} and Sukstanskii {\it et al.}~\cite{JAP89p775} investigated the MI behavior in multilayers, analyzing the influence of width, length, and relative conductivity in  MI effect.

% while Antonov {\it et al.}~\cite{JPDAP32p1204} showed the effect of non-magnetic metal thickness on the MI curves.

% {\color{blue} Anyway, for all cases, the theoretical studies have been mainly focused in the transverse magnetic permeability and in a limited low frequency range.}

%{\color{red} Generally, the key to understanding and controlling MI lies in the knowledge of transverse magnetic permeability behavior in a given material. Although the magnetoimpedance effect has been extensively investigated, many questions still remain unclear. Unquestionably, the aforementioned models represent a major breakthrough to the understanding of the magnetoimpedance effect. However, although the MI results obtained are consistent and seem to reproduce experimental data, the main question concerns knowledge of transverse permeability behavior over a wide frequency range. This restricts the results to a limited frequency range or hinders identification of which permeability model must be chosen for that limited range.

%Since experimental measurements are usually taken over a wide range of frequencies, in which different mechanisms contributes to the permeability and MI variations, a general theoretical approach to the transverse magnetic permeability which enables the MI calculation, considering frequency dependent magnetic permeability, becomes very important for MI interpretation.}

Although the MI results obtained are consistent and seem to reproduce experimental data, they are restricted to a limited frequency range. Since experimental measurements are usually taken over a wide range of frequencies, in which different mechanisms contributes to the permeability, a general theoretical approach to the transverse magnetic permeability which enables the MI calculation, considering frequency dependent magnetic permeability, becomes very important for MI interpretation.

In this paper, we report a theoretical and experimental investigation of the magnetoimpedance effect for the thin film geometry in a wide frequency range. First of all, we perform numerical calculations of the longitudinal magnetoimpedance for single layered, multilayered and exchange bias systems, from a classical electromagnetic impedance for a planar system. To this end, we consider a theoretical approach that takes into account a magnetic permeability model for planar geometry and the appropriate magnetic free energy density for each structure. We analyze the magnetoimpedance behavior, and discuss the main features and advantages of each structure, as well as we relate the numerical calculations with experimental results found in literature. Finally, to test the robustness of the approach, we compare theoretical results calculated for an exchange biased multilayered system with experimental magnetoimpedance measurements obtained in a wide range of frequencies for an exchange biased multilayered film. Thus, we provide experimental evidence to confirm the validity of the theoretical approach to describe the magnetoimpedance in ferromagnetic films.

\section{Theoretical approach}
\label{Theoretical_approach}

\subsection{Thin film planar geometry}

\label{Numerical_calculation_theoretical_system}

To investigate the MI effect, we perform numerical calculations of quasi-static magnetization curves, magnetic permeability and magnetoimpedance for the thin film geometry. To this end, from the appropriate magnetic free energy density for the investigated structure, in a first moment, we consider a general magnetic susceptibility model which takes into account its dependence with both frequency and magnetic field~\cite{JMMM296p1}. It is therefore possible to obtain the transverse magnetic permeability for planar geometry from susceptibility and in turn describe the MI behavior by using different models, according to system structure, for a wide range of frequencies and external magnetic fields.

\begin{figure}[!]
\includegraphics[width=8.5cm]{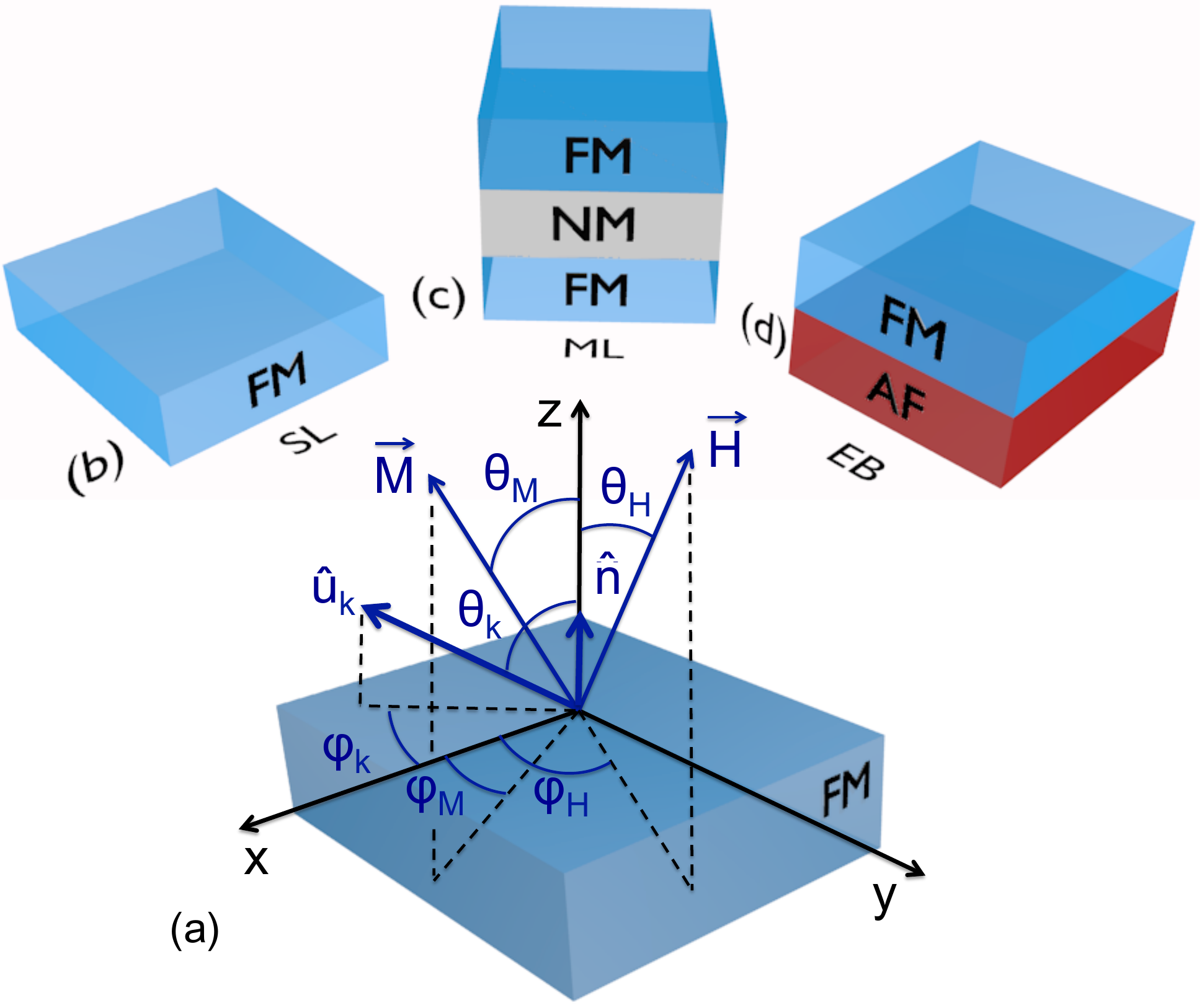}
\caption {Ferromagnetic thin films modeled as a planar system. (a) Schematic diagram of the theoretical ferromagnetic system and definitions of magnetization and magnetic field vectors considered for the numerical calculation of magnetization, magnetic permeability, and magnetoimpedance curves. We consider $\vec H$ as the external magnetic field vector, defined by the angles $\theta_H$ and $\varphi_H$ that describe the orientation of the field vector, $\vec M$ as the magnetization vector, with $\theta_M$ and $\varphi_M$ as the equilibrium angles of magnetization for a given magnetic field value, $\hat u_k$ as the unit vector along the uniaxial anisotropy direction, defined by $\theta_k$ and $\varphi_k$, and $\hat n$ as a unit vector normal to the film plane. 
For the numerical calculations, the external magnetic field and electrical current are in the film plane along the $y$-direction, and, due to the planar configuration, the transverse magnetic permeability is considered along the $x$-direction. (b) Single layered (SL) system, composed by a $500$ nm-thick ferromagnetic (FM) layer. (c) Multilayered (ML) system, composed by $250$ nm-thick ferromagnetic layers and metallic non-magnetic (NM) layers with variable thicknesses. (d) Exchange-biased (EB) system, composed by a single $500$ nm-thick ferromagnetic layer and a single antiferromagnetic (AF) layer.
%For the numerical calculations, the external magnetic field and electrical current are in the film plane along the $y$-direction, and, due to the planar configuration, the transverse magnetic permeability is considered along the $x$-direction. (b) Single layered (SL) system, composed by a ferromagnetic (FM) $500$ nm-thick NiFe film. (c) Multilayered (ML) system, composed by $250$ nm-thick NiFe ferromagnetic layers and metallic Ag non-magnetic (NM) layers with variable thicknesses. (d) Exchange-biased (EB) system, composed by a single ferromagnetic $500$ nm-thick NiFe film and a single antiferromagnetic (AF) IrMn layer.
}
\label{Fig_01}
\end{figure}

We focus on the study of ferromagnetic thin films, which can be modeled as a planar system. Here, in particular, we calculate the longitudinal magnetoimpedance effect for single layered, multilayered or exchange biased systems. Figure~\ref{Fig_01}(a) presents the theoretical system and the definitions of the relevant vectors considered to perform the numerical calculations. In order to investigate the magnetoimpedance effect in films, we consider the single layered, multilayered and exchange biased systems, as respectively shown in Fig.~\ref{Fig_01}(b)-(d). 

Thus, from the appropriate magnetic free energy density $\xi$ for each structure, a routine for energy minimization determine the values the equilibrium angles $\theta_M$ and $\varphi_M$ of magnetization for a given external magnetic field $\vec H$, and we obtain the magnetization curve, permeability tensor $\mu$ and longitudinal magnetoimpedance $Z$ for the respective structure in a wide range of frequencies.

\subsection{Permeability Tensor}
\label{Numerical_calculation_permeability_tensor}

Generally, the magnetization dynamics is governed by the Landau-Lifshitz-Gilbert equation, given by
\begin{equation}
\frac{d\vec{M}}{dt}=-\gamma (\vec{M}\times \vec{H_{eff}}) - \gamma \frac{\alpha}{M} \left[\vec{M} \times (\vec{M} \times \vec{H_{eff}}) \right],
\label{llg}
\end{equation} 
\noindent where $\vec M$ is the magnetization vector, $\vec H_{eff}$ is the effective magnetic field, and $\gamma = |\gamma_{G}|/(1+\alpha^2)$, in which $\gamma_{G}$ is the gyromagnetic ratio and $\alpha$ the phenomenological Gilbert damping constant. In a MI experiment, the effective magnetic field presents two contributions and can be written as $\vec H_{eff} = (\vec H + \vec H_\xi) + \vec h_{ac}$. The first term, $(\vec H + \vec H_\xi)$, corresponds to the static component of the field. It contains the external magnetic field $\vec H$ and the internal magnetic field $\vec{H}_{\xi} = -\frac{\partial \xi} {\partial \vec M}$~\cite{PRR10p113}, due to different contributions to the magnetic free energy density $\xi$, such as magnetic anisotropies and induced internal magnetic fields. On the other hand, the second term corresponds to the alternate magnetic field $\vec h_{ac}$ generated by the $I_{ac}$ applied to the sample, which in turn induces deviations of the magnetization vector from the static equilibrium position. Equation~(\ref{llg}) is a general expression that can be applied to express the magnetization dynamics of any system, with any geometry.
%Thereby, applying the drive current and $\vec{H}$ field in the same direction, the fields configurations lead to a similar arrangement to that used in FMR experiments \cite{TPPM}. 

As previously cited, it is possible to understand the MI effect from the knowledge of the transverse magnetic permeability of a given material. This goal is achieved by considering how magnetic dynamics transition takes place from one state of equilibrium to another under both {\it dc} and {\it ac} fields. With this spirit, a very interesting approach to study the magnetization dynamics was successfully undertaken by Spinu {\it et al.}~\cite{JMMM296p1}. This theory allows us to investigate the magnetic susceptibility tensor and its dependence on both frequency and magnetic field, using knowledge of appropriate magnetic free energy density. 

From the approach~\cite{JMMM296p1}, the magnetic susceptibility tensor, in spherical coordinates, for a general system with a given magnetic free energy density $\xi$, is written as
\begin{multline} \label{tsus}
\begin{split}
\chi (r, \theta, \varphi) = \eta \gamma^2 (1+\alpha^2)   \left( \begin{array}{ccc}
0 & 0 & 0 \\
0 & \frac{ \xi_{\varphi \varphi}}{\sin^2 \theta_{M}} & -\frac{ \xi_{\theta \varphi}}{\sin \theta_{M}}\\
0 &- \frac{  \xi_{\theta \varphi}}{\sin \theta_{M}} &  \xi_{\theta \theta} \end{array} \right) \\  + \eta \left( \begin{array}{ccc}
0 & 0 & 0 \\
0 & i  M_{s} \gamma \omega \alpha & i  M_{s} \gamma \omega\\
0 & -i  M_{s} \gamma \omega & i  M_{s} \gamma \omega \alpha \end{array} \right),
\end{split}
\end{multline}
\noindent where $\eta$ is
\begin{equation}
\eta = \frac{1}{\omega^{2}_{r} - \omega^2 + i \omega \Delta \omega}.
\label{FMR}
\end{equation}
\noindent The quantities $\omega_{r}$ and $\Delta \omega$ in Eq.~(\ref{FMR}) are known, respectively, as the resonance frequency and width of the resonance absorption line, given by~\cite{TPPM,JMMM296p1}
\begin{equation}
\omega_{r} = \frac{\gamma}{M \sin \theta_{M}} \sqrt{1+\alpha^2} \sqrt{\xi_{\theta \theta} \xi_{\varphi \varphi} - \xi^{2}_{\theta \varphi}}, 
\end{equation}
\noindent and
\begin{equation}
 \Delta \omega = \frac{\alpha \gamma}{M} \left( \xi_{\theta \theta} + \frac{\xi_{\varphi \varphi}}{\sin^{2} \theta_{M}} \right).
 \end{equation}
\noindent Here, $\xi_{\theta \theta}$, $\xi_{\varphi \varphi}$, $\xi_{\varphi \theta}$, and $\xi_{\theta \varphi}$ are the second derivatives of the magnetic free energy density at an equilibrium position, defined by the magnetization vector with $\theta_{M}$ and $\varphi_{M}$, as previously shown in Fig.~\ref{Fig_01}(a).

Considering the matrix of the linear transformation of the unit vectors from spherical to Cartesian coordinates, the susceptibility tensor in the laboratory reference can be obtained. For instance, the real and imaginary components of the term $\chi_{xx}$ can be, respectively, written as~\cite{JMMM296p1}
\begin{widetext} 
\begin{equation} \small \label{sure}
\begin{aligned}
\Re[\chi_{xx}] =\kappa \left[ 
\begin{array}{c}
\left(\alpha ^2+1\right) \gamma  \omega _r^2
   \left( \xi_{\varphi \varphi}\cot ^2\theta_M \,\cos ^2\varphi_M\,-2\xi_{\theta \varphi}\cot
   \theta_M \, \sin \varphi_M \, \cos \varphi_M \,
   + {\xi_{\theta \theta}}\sin ^2\varphi_M\right) \\
 -\omega ^2 \left[  \begin{array}{c}-2\left(\alpha ^2+1\right) \gamma 
   \xi_{\theta \varphi}\cot \theta_M \, \sin \varphi_M \, \cos \varphi_M \,
    \\
   +\left(\left(\alpha ^2+1\right) \gamma  {\frac{ \xi_{\varphi \varphi}}{\sin^2 \theta_{M}}}-\alpha  M_s
   \triangle \omega \right)\cos ^2 \theta_M \, \cos ^2\varphi_M \,
   +\left(\left(\alpha
   ^2+1\right) \gamma  {\xi_{\theta \theta}}-\alpha  M_s \triangle \omega
   \right)\sin ^2 \varphi_M \, \end{array}\right]
\end{array} 
\right],
\end{aligned}
\end{equation}  
\end{widetext}
   \begin{widetext} 
\begin{equation} \small \label{suim}
\begin{aligned}
\Im[\chi_{xx}] = -\kappa  \omega \left[ 
\begin{array}{c}
-2\xi_{\theta \varphi}\left(\alpha ^2+1\right) \gamma 
   \triangle \omega  \cot \theta_M \,\sin \varphi_M \,\cos
   \varphi_M \,+\left(\left(\alpha ^2+1\right) \gamma 
    \triangle \omega {\frac{ \xi_{\varphi \varphi}}{\sin^2 \theta_{M}}} +\alpha  M_s \omega ^2\right)\cos ^2\theta_M \,\cos
   ^2\varphi_M \, \\
+ \left(\left(\alpha ^2+1\right) \gamma 
   \triangle \omega  {\xi_{\theta \theta}} +\alpha  M_s \omega
   ^2\right)\sin^2\varphi_M -\alpha  M_s \omega _r^2 \left(\cos ^2\theta_M \,\cos
   ^2\varphi_M \,+\sin ^2\varphi_M\right)
\end{array} 
\right],
\end{aligned}
\end{equation}
\end{widetext}
\noindent where 
\[ \kappa = \frac{\gamma}{(\omega_{r}^{2}-\omega^{2})^{2}+\omega^2 \Delta \omega^2}.\]

In particular, the diagonal component of the susceptibilty tensor presented in Eqs.~(\ref{sure})-(\ref{suim}), as well as the $\chi_{yy}$ and $\chi_{zz}$ components (not presented here for sake of simplicity), exhibit form similar to that presented in Ref.~\cite{JMMM296p1} when $\omega \rightarrow 0$, as expected. From the cited equations, it can be noticed a clear dependence of the magnetic susceptibility with the equilibrium angles of the magnetization, as well as with the derivatives of the magnetic free energy density. Thus, this general description to the susceptibility and, consequently, to the dynamic magnetic behavior corresponds to a powerful tool, once it can be employed for any magnetic structure, using an appropriate energy configuration. 

In ferromagnetic thin films, which can be modeled as planar systems, the magnetization is frequently observed to be in the plane of the film. Thus, by considering $\theta_{M} = 90^{\circ}$ (See Fig.~\ref{Fig_01}(a)), the expressions for the terms of the permeability tensor $\mu = 1+4 \pi \chi$ can be considerably simplified. The diagonal terms represented by $\mu_{xx}$, $\mu_{yy}$ and $\mu_{zz}$ can be written as
\begin{equation} \small \label{muxx}
\begin{aligned}
& \mu_{xx} =  1 + 4 \pi  \kappa  \sin^2 \varphi_{M} \\ 
 & \times \left[ \begin{array}{c}   (\omega^2_r -\omega^2) (1+\alpha^2) \gamma \xi_{\theta \theta} + \alpha M_{s} \omega^2 \triangle \omega  \\
  +  i  \left[-\left(1+\alpha ^2\right)\gamma  \omega \triangle \omega  \xi _{\theta \theta } + \alpha  M_{s} \omega \left(\omega _r^2-\omega^2\right)\right] \end{array} \right],
\end{aligned}
\end{equation}

\begin{equation} \small \label{muyy}
\begin{aligned}
& \mu_{yy} =1+  4 \pi  \kappa  \cos^2 \varphi_{M}  \\
& \times  \left[ \begin{array}{c}  (\omega^2_r -\omega^2) (1+\alpha^2) \gamma \xi_{\theta \theta} + \alpha M_{s} \omega^2 \Delta \omega  \\
+ i  \left[-\left(1+ \alpha ^2\right) \gamma \omega \triangle \omega  \xi _{\theta \theta }+ \alpha  M_{s} \omega \left(\omega _r^2-\omega^2\right) \right]\end{array} \right],
  \end{aligned}
\end{equation}

\begin{equation} \small  \label{muzz}
\begin{aligned}
& \mu_{zz} = 1 + 4 \pi  \kappa   \\
&\times \left[ \begin{array}{c} \left(\omega^2-\omega^2_r\right) \left(1+ \alpha ^2 \right) \gamma  \xi _{\varphi \varphi }  + \alpha M_{s} \omega ^2 \triangle \omega \\ 

+ i   \left[-\left(1+ \alpha ^2\right) \gamma \omega  \triangle \omega  \xi _{\varphi \varphi }+ \alpha M_{s} \omega  \left(\omega _r^2-\omega^2\right)\right] \end{array} \right].
\end{aligned}
\end{equation}

Moreover, the off-diagonal terms are
\begin{equation} \small  \label{muxy}
\begin{aligned}
& \mu_{xy} = \mu_{yx} = 1 + 2 \pi  \kappa  \sin (2 \varphi_{M} ) \\
& \times \left[ \begin{array}{c} - \left(\omega _r^2-\omega ^2\right)  \left(1+\alpha ^2\right) \gamma \xi _{\theta \theta }  - \alpha  M_{s} \omega^2 \triangle \omega \\
+ i \left[\left(1+ \alpha ^2\right) \gamma \omega \triangle \omega  \xi _{\theta \theta }- \alpha  M_{s} \omega \left(\omega _r^2-\omega^2\right)\right] \end{array} \right],
  \end{aligned}
\end{equation}

\begin{equation} \small  \label{muxz}
\begin{aligned}
& \mu_{xz} = 1 + 4 \pi  \kappa  \sin \varphi_{M}  \\
& \times \left[  \begin{array}{c} - \left(\omega _r^2-\omega^2\right) \left(1+\alpha ^2\right) \gamma  \xi _{\theta \varphi } - M_{s} \omega ^2 \triangle \omega \\
+ i  \left[\left(1+\alpha ^2\right) \gamma \omega \triangle \omega  \xi _{\theta \varphi }+ M_{s} \omega \left(\omega _r^2-\omega ^2\right)\right] \end{array} \right],
  \end{aligned}
\end{equation}

\begin{equation} \small  \label{muyz}
\begin{aligned}
& \mu_{yz} = 1 + 4 \pi  \kappa  \cos \varphi_{M}  \\
& \times \left[  \begin{array}{c} \left(\omega _r^2-\omega^2\right) \left(1+\alpha ^2\right) \gamma  \xi _{\theta \varphi } + M_{s} \omega ^2 \triangle \omega \\
+ i  \left[-\left(1+\alpha ^2\right) \gamma \omega \triangle \omega  \xi _{\theta \varphi } - M_{s} \omega \left(\omega _r^2-\omega ^2\right)\right] \end{array} \right],
  \end{aligned}
\end{equation}

\begin{equation} \small  \label{muzx}
\begin{aligned}
& \mu_{zx}  = 1 + 4 \pi  \kappa  \sin \varphi_{M}  \\
& \times \left[  \begin{array}{c} - \left(\omega _r^2-\omega^2\right) \left(1+\alpha ^2\right) \gamma  \xi _{\theta \varphi } + M_{s} \omega ^2 \triangle \omega \\
+ i  \left[\left(1+\alpha ^2\right) \gamma \omega \triangle \omega  \xi _{\theta \varphi }+ M_{s} \omega \left(\omega _r^2-\omega ^2\right)\right] \end{array} \right],
     \end{aligned}
\end{equation}

\begin{equation} \small  \label{muzy}
\begin{aligned}
& \mu_{zy} = 1 + 4 \pi  \kappa  \cos \varphi_{M}  \\
& \times \left[  \begin{array}{c} \left(\omega _r^2-\omega^2\right) \left(1+\alpha ^2\right) \gamma  \xi _{\theta \varphi } - M_{s} \omega ^2 \triangle \omega \\
+ i  \left[-\left(1+\alpha ^2\right) \gamma \omega \triangle \omega  \xi _{\theta \varphi } - M_{s} \omega \left(\omega _r^2-\omega ^2\right)\right] \end{array} \right].
     \end{aligned}
\end{equation}

%From this permeability tensor, it is possible to calculate the magnetoimpendance effect for any planar structure, since an appropriate magnetic free energy density is considered. 

For all the numerical calculations, we consider that the magnetic field $\vec H$ is applied, as well as the electrical curren is flowing, along the $y$-direction (See Fig.~\ref{Fig_01}(a)), $\theta_H = \varphi_H = 90^\circ$. Then, the $\mu_{xx}$ term can be understood as the transverse magnetic permeability $\mu_t$. In the next sections, we present the MI calculations for single layered, multilayered and exchange biased systems. To this end, we consider MI models, such as the classical MI expression for a slab conductor~\cite{LLmedia, JMMM195p764} or the model proposed by Panina for multilayers~\cite{SAAP81p71}, previously explored in a limited frequency range. In particular, this limitation is due to the employed permeability calculation. Here, we consider a general approach to the permeability and, consequently, we are able to explore the MI behavior in several planar structures in a wide frequency range.

\subsection{Single layered system}
\label{Numerical_calculation_single_layer}

First of all, we perform numerical calculation for the longitudinal MI effect for a single layered system, as presented in Fig.~\ref{Fig_01}(b). 

We consider a Stoner-Wohlfarth modified model to describe the magnetic free energy density. In this case, it can be written as
\begin{equation}
\label{esl}
\xi = - \vec{M} \cdot \vec H - \frac{H_{k}}{2 M_s} \left(  \vec{M} \cdot \hat{u}_{k} \right)^2+4 \pi M^{2}_{s} \left( \hat{M} \cdot \hat{n} \right),
\end{equation}
\noindent where the first term is the Zeeman interaction, the second term describes the uniaxial anisotropy and the third one corresponds to the demagnetizing energy density for a thin planar system, such as a thin film. In this case, in addition to the vectors $\vec H$, $\vec M$, $\hat u_{k}$, and $\hat n$ already discussed in Fig.~\ref{Fig_01}(a), $H_{k}=2K_{u}/M_{s}$ is the known anisotropy field, $K_u$ is the uniaxial anisotropy constant, and $M_s$ is the saturation magnetization of the ferromagnetic material.

The longitudinal impedance is strongly dependent of the sample geometry. Here, to describe the magnetoimpedance in a single layered system, we consider the approach reported by Kraus~\cite{JMMM195p764} for an infinite slab magnetic conductor. Thus, for a single layered system, the impedance can be written as~\cite{JMMM195p764}
\begin{equation}
\label{sl}
\frac{Z}{R_{dc}} = k \frac{t}{2} \coth{\left(  k \frac{t}{2} \right)},
\end{equation}
\noindent where $R_{dc}$ is the electrical dc resistance, $t$ is the thickness of the system, and $k=(1-i)/\delta$, where $\delta$ is the classic skin depth, given by
\begin{equation}
\label{sd}
\delta= \sqrt{2 \rho/ \omega \mu}\,,
\end{equation}
\noindent in which $\rho$ is the electric resistivity, $\omega$ is the angular frequency, and $\mu$ is the magnetic permeability. In our case, we consider $\mu = \mu_{xx} = \mu_t$.

Thus, from the magnetic free energy density, given by Eq.~(\ref{esl}), and the calculation of the transverse magnetic permeability, Eq.~(\ref{muxx}), the longitudinal magnetoimpedance for a single layered system, Eq.~(\ref{sl}), can be obtained. The other terms of the permeability tensor previously presented can be used to calculate the $Z$ behavior, since a specific calculation of the $Z$ tensor is done. 

\begin{figure}[!ht]
\includegraphics[width=8.5cm]{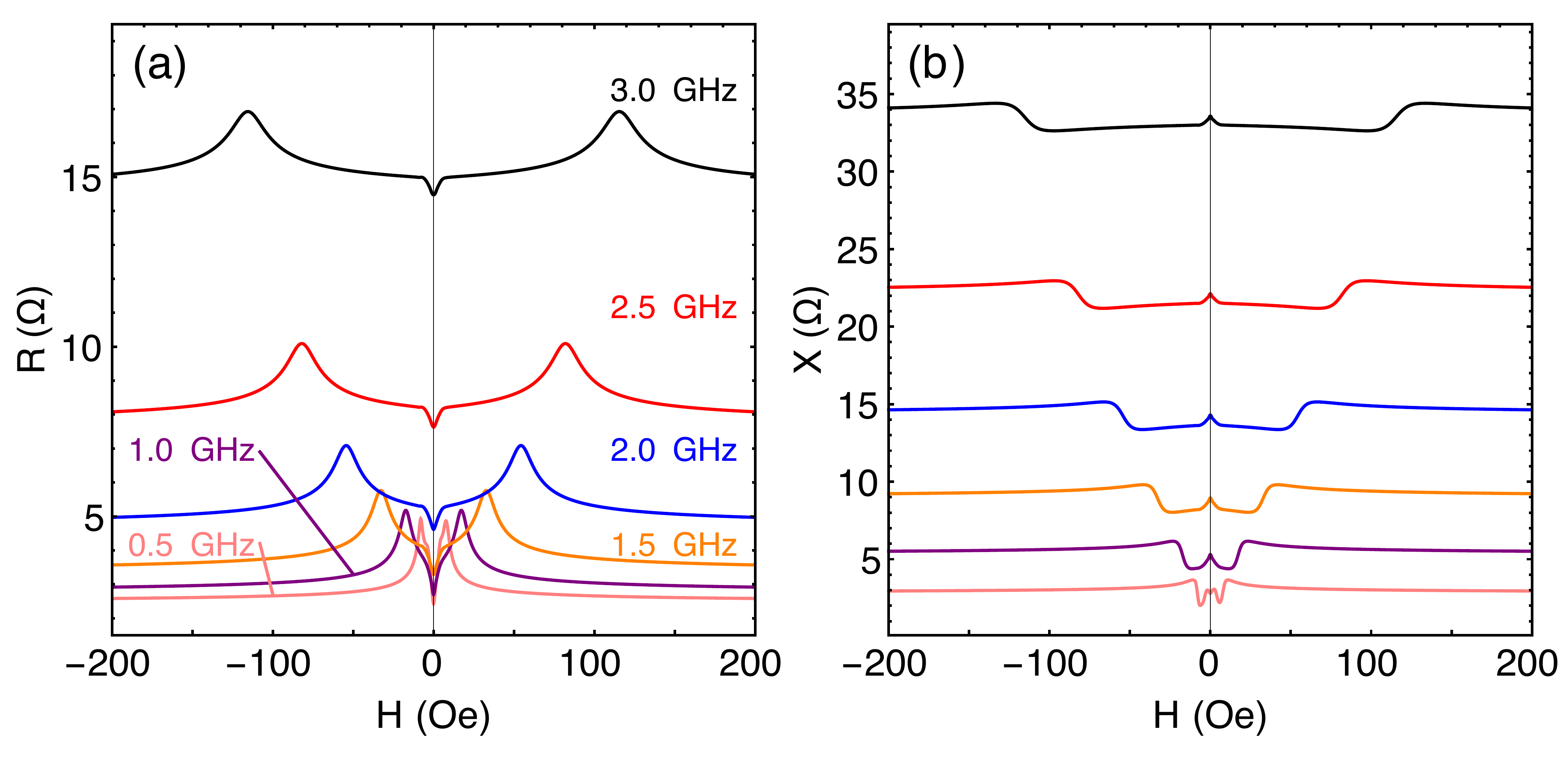}
\caption {(a) Real $R$ and (b) imaginary $X$ components of the longitudinal impedance as a function of the external magnetic field for selected frequency values. The numerical calculations are obtained, using the described approach, for a single layered system with $M_s = 780 $ emu/cm$^3$, $H_{k} = 5$ Oe, $\theta_{k} = 90^{\circ}$, $\varphi_{k} = 2^{\circ}$, $\alpha = 0.018$, $\gamma_G / 2 \pi = 2.9$ MHz/Oe~\cite{JAP100p023906}, $t=500$ nm, $\theta_{H} = 90^{\circ}$, and $\varphi_{H} = 90^{\circ}$.}
\label{Fig_02}
\end{figure}

For a single layered system, to perform the numerical calculation, we consider the following parameters: $M_s = 780 $ emu/cm$^3$, $H_{k} = 5$ Oe, $\theta_{k} = 90^{\circ}$, $\varphi_{k} = 2^{\circ}$, $\alpha = 0.018$, $\gamma_G / 2 \pi = 2.9$ MHz/Oe \cite{JAP100p023906}, $t=500$ nm, $\theta_{H} = 90^{\circ}$, and $\varphi_{H} = 90^{\circ}$. We intentionaly chose $\varphi_k \neq 0^\circ$ since small deviations in the sample position or of the magnetic field in an experiment are reasonable. Figure~\ref{Fig_02} shows the numerical calculations for the real $R$ and imaginary $X$ components of the longitudinal impedance as a function of the external magnetic field for selected frequency values.

It is important to point out that, experimentally, the MI measurements presents a frequency dependent shift of the real and imaginary components, a feature related to the electrical/metallic contributions of the sample and of the microwave cavity or microstrip employed in the experiment. In order to directly compare experimental data with numerical calculation, this dependence can be removed of the experimental MI results, according to Ref.~\cite{JAP99p08C507}, or can be inserted in the MI numerical calculation by fitting the measured $R$ and $X$ curves as a function of the frequency for the highest magnetic field value~\cite{JAP110p093914}, where the sample is magnetically saturated. In this case of a single layered system, we consider a fitting obtained from the data reported in Ref.~\cite{JPDAP43p295004}.

Thus, from Fig.~\ref{Fig_02}, the well-known symmetric magnetoimpedance behavior around $H=0$ for anisotropic systems is verified, including the dependence with the magnetic field amplitude, frequency, and the orientation between the applied magnetic field and {\it ac} current with respect to the magnetic anisotropies. A double peak behavior is present for the whole frequency range, a feature of the FMR relation dispersion~\cite{JAP110p093914,SAAP106p187,JMMM195p764}, in a signature of the parallel alignment of the external magnetic field and {\it ac} current along the hard magnetization axis. 

At low and intermediate frequencies (not shown), below $0.5$ GHz, the position of the peaks remains unchanged and they are close to $H_{k}$. This feature reflects the fact that, at this frequency range, the skin effect is the main responsible for the magnetization dynamics and MI variations. Beyond $0.5$ GHz, besides the skin effect, the FMR effect also becomes an important mechanism responsible for variations in MI effect, a fact evidenced by the displacement of the peak position in the double peak structure toward higher fields as the frequency is increased following the the behavior predicted for the FMR effect~\cite{JAP110p093914,SAAP106p187,JMMM195p764}. The contribution of the FMR effect to $Z$ is also verified using the method described by Barandiar\'{a}n {\it et al.}~\cite{JAP99p103904}, and previously employed by our group in~\cite{JPDAP43p295004}. In particular, the classical FMR signature is observed in the numerical calculation of the longitudinal MI response at the high frequency range strictly due to the fact that we employ a magnetic permeability model derived from the FMR theory~\cite{JMMM296p1}. 

These numerical calculation results are in qualitative agreement with several experimental results for single layered thin films~\cite{APL67p3346,JAP101p033908,EFSilva_unpublished} with uniaxial magnetic behavior, when the magnetic field and current are transverse to the easy magnetization axis during the experiment.

\subsection{Multilayered system}
\label{Numerical_calculation_multilayer}

Here, we perform the numerical calculation of the longitudinal MI effect for a multilayered system, as presented in Fig.~\ref{Fig_01}(c).

The multilayered system consists of two ferromagnetic layers separated by a metallic non-magnetic layer. To model it, we consider a Stoner-Wohlfarth modified model, similar to that discussed in Subsection~\ref{Numerical_calculation_single_layer}, and the magnetic free energy density can be written as
\begin{equation}
\label{eml}
\xi = \sum_{i=1}^{2} \left[ - \vec{M}_i \cdot \vec H - \frac{H_{ki}}{2 M_{si}} \left(  \vec{M}_{i}\cdot \hat{u}_{ki} \right)^2+4 \pi M^{2}_{si} \left( \hat{M}_{i} \cdot \hat{n} \right) \right],
\end{equation}  
\noindent where $\vec M_{si}$ and $M_{si}$ are the magnetization vector and saturation magnetization for each ferromagnetic layer, respectively, $H_{ki} = 2K_{ui}/M_{si}$ is the anisotropy field for each layer, and $K_{ui}$ is the uniaxial anisotropy constant, directed along $\hat u_{ki}$, for each layer. In a traditional multilayered system, it is reasonable to consider $M_{s1} = M_{s2} = M_{s}$, $K_{u1}= K_{u2}= K_{u}$, $\hat u_{k1}=\hat u_{k2}=\hat u_{k}$, since the two layers are made of similar ferromagnets.

To describe the magnetoimpedance behavior in a multilayered system, we consider the approach to study the magnetoimpedance effect in a trilayered system reported by Panina {\it et al.}\cite{SAAP81p71} and investigated by our group~\cite{JAP110p093914}. In this model, the trilayered system has finite width $2b$ and length $l$ for all layers, thicknesses $t_1$ and $t_2$, and conductivity values $\sigma_1$ and $\sigma_2$ for the metallic non-magnetic and ferromagnetic layers, respectively, and variable flux leaks across the inner non-magnetic conductor. When $b$ is sufficiently large and the edge effect is neglected, impedance is dependent on the film thickness $t$. Therefore, for a tri-layered system, impedance can be written as
\begin{equation}
\label{impmult}
\frac{Z}{R_{dc}} = (\eta_m \eta_f)  \left[  \frac{\coth{\left(\frac{\eta_m \sigma_2}{\mu \sigma_1 } \right)} \coth {(\eta_f)} +\frac{2 \eta_m}{k_1 t_1}}{\coth{\left(  \frac{\eta_m \sigma_2}{\mu\sigma_1 }  \right) +\frac{2 \eta_m}{k_1 t_1} \coth{(\eta_f)} } }  \right],
\end{equation}
\noindent where $\mu$ is the magnetic permeability for the ferromagnetic layers, in our case, we consider $\mu = \mu_{xx} = \mu_t$, and 
\[ \eta_m = \frac{k_1 t_1}{2} \left( \frac{\mu\sigma_1  }{\sigma_2} \right),  \hspace{0.5cm} \eta_f = k_2 t_2, \]
\[  k_1 = \frac{(1-i)}{\delta_1}, \hspace{0.5 cm} k_2 = \frac{(1-i)}{\delta_2}, \]
\[  \delta_1 = (2 \pi \sigma_1 \omega)^{-1/2} , \hspace{0.5 cm} \delta_2= ( 2 \pi \sigma_2 \omega \mu)^{-1/2}. \]

To perform the numerical calculation for a multilayered system, we consider the parameters similar to those previously employed: $M_{s1}=M_{s2} = 780$ em/cm$^3$ , $H_{k1}=H_{k2} = 5$ Oe, $\theta_{k1}=\theta_{k2} = 90^{\circ}$, $\varphi_{k1} = \varphi_{k2} = 2^{\circ}$, $\alpha = 0.018$, $\gamma_G / 2 \pi = 2.9$ MHz/Oe, $t_{1} = 100$ nm, $t_{2} = 250$ nm, $\theta_{H} = 90^{\circ}$, and $\varphi_{H} = 90^{\circ}$. In particular, the thickness of the metallic non-magnetic layer is thick enough to neglect the bilinear and biquadratic coupling between the ferromagnetic layers. Moreover, we employ $\sigma_1 = 6 \times 10^7\,(\Omega\textrm{m})^{-1}$, and $\sigma_2 = \sigma_1/4$. 

\begin{figure}[!]
\includegraphics[width=8.5cm]{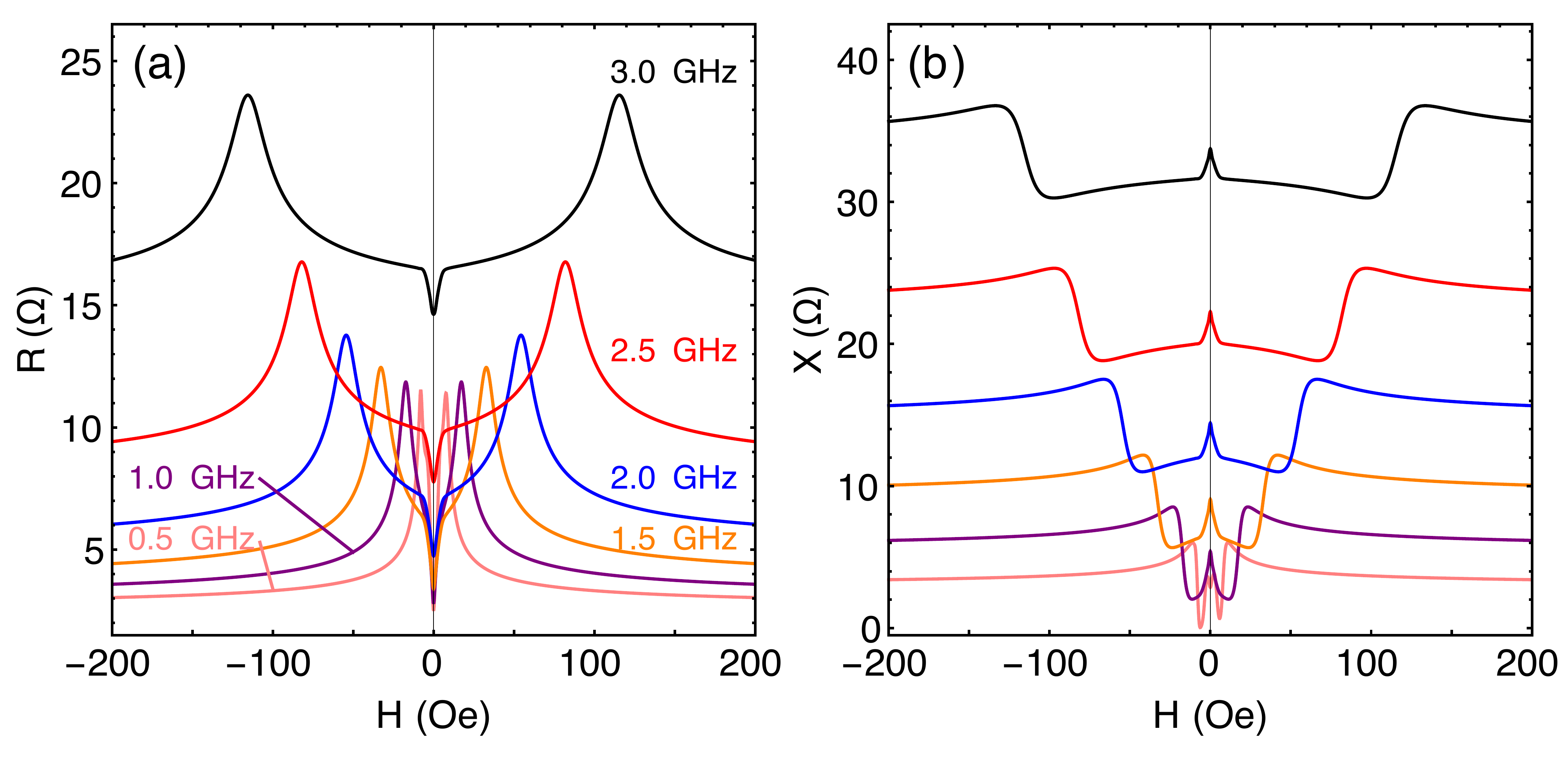}
\caption {(a) Real $R$ and (b) imaginary $X$ components of the longitudinal impedance as a function of the external magnetic field for selected frequency values. The numerical calculations are obtained for a multilayered system with $M_{s1}=M_{s2} = 780$ em/cm$^3$ , $H_{k1}=H_{k2} = 5$ Oe, $\theta_{k1}=\theta_{k2} = 90^{\circ}$, $\varphi_{k1} = \varphi_{k2} = 2^{\circ}$, $\alpha = 0.018$, $\gamma_G / 2 \pi = 2.9$ MHz/Oe, $t_{1} = 100$ nm, $t_{2} = 250$ nm, $\sigma_1 = 6 \times 10^7\,(\Omega\textrm{m})^{-1}$, $\sigma_2 = \sigma_1/4$, $\theta_{H} = 90^{\circ}$, and $\varphi_{H} = 90^{\circ}$.}
\label{Fig_03}
\end{figure}

Thus, from Eqs.~(\ref{eml}), (\ref{muxx}) and (\ref{impmult}), Fig.~\ref{Fig_03} shows the numerical calculations for the real $R$ and imaginary $X$ components of the longitudinal impedance as a function of the external magnetic field for selected frequency values. In particular, for a multilayered system, the MI curves present all the typical features described for an anisotropic single layered system, including the double peak MI structure due to the orientation between $\vec H$, $I_{ac}$ sense and $\hat u_k$, as well as the $R$, $X$ and $Z$ behavior with frequency. In order to consider the frequency dependent shift of $R$ and $X$ for a multilayered system, we consider a fitting obtained from the data reported in Ref.~\cite{JPDAP43p295004}.

\begin{figure}[!]
\includegraphics[width=8.5cm]{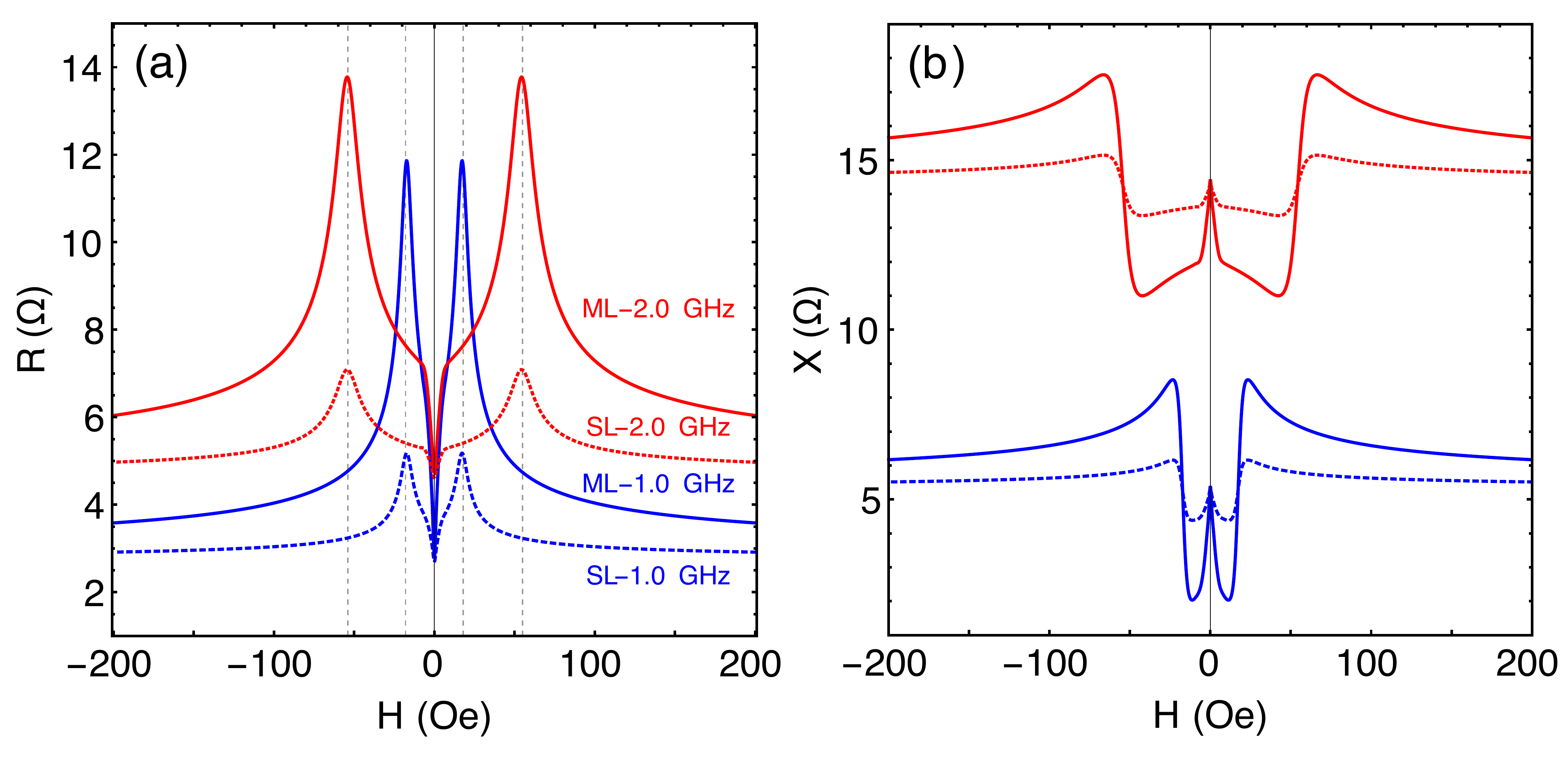}
\caption{Comparision of the (a) real $R$ and (b) imaginary $X$ components of the longitudinal impedance, as a function of the external magnetic field for selected frequency values, calculated for single layered (Dashed lines) and multilayered (Solid lines) systems. The numerical calculations are performed using parameters similar to those previously employed for single layered and multilayered systems.}
\label{Fig_04}
\end{figure}

To contrast the MI behavior verified for the studied systems, Fig.~\ref{Fig_04} present a comparison between the numerical calculations of the real $R$ and imaginary $X$ components of the longitudinal impedance for single layered and multilayered systems. The positions in field of the MI peaks are similar, irrespective of the frequency. This feature is expected in this case, once similar parameter values are employed for both numerical calculations, and, consequently, both systems have the same quasi-static magnetic properties. 

\begin{figure}[!]
\includegraphics[width=8.5cm]{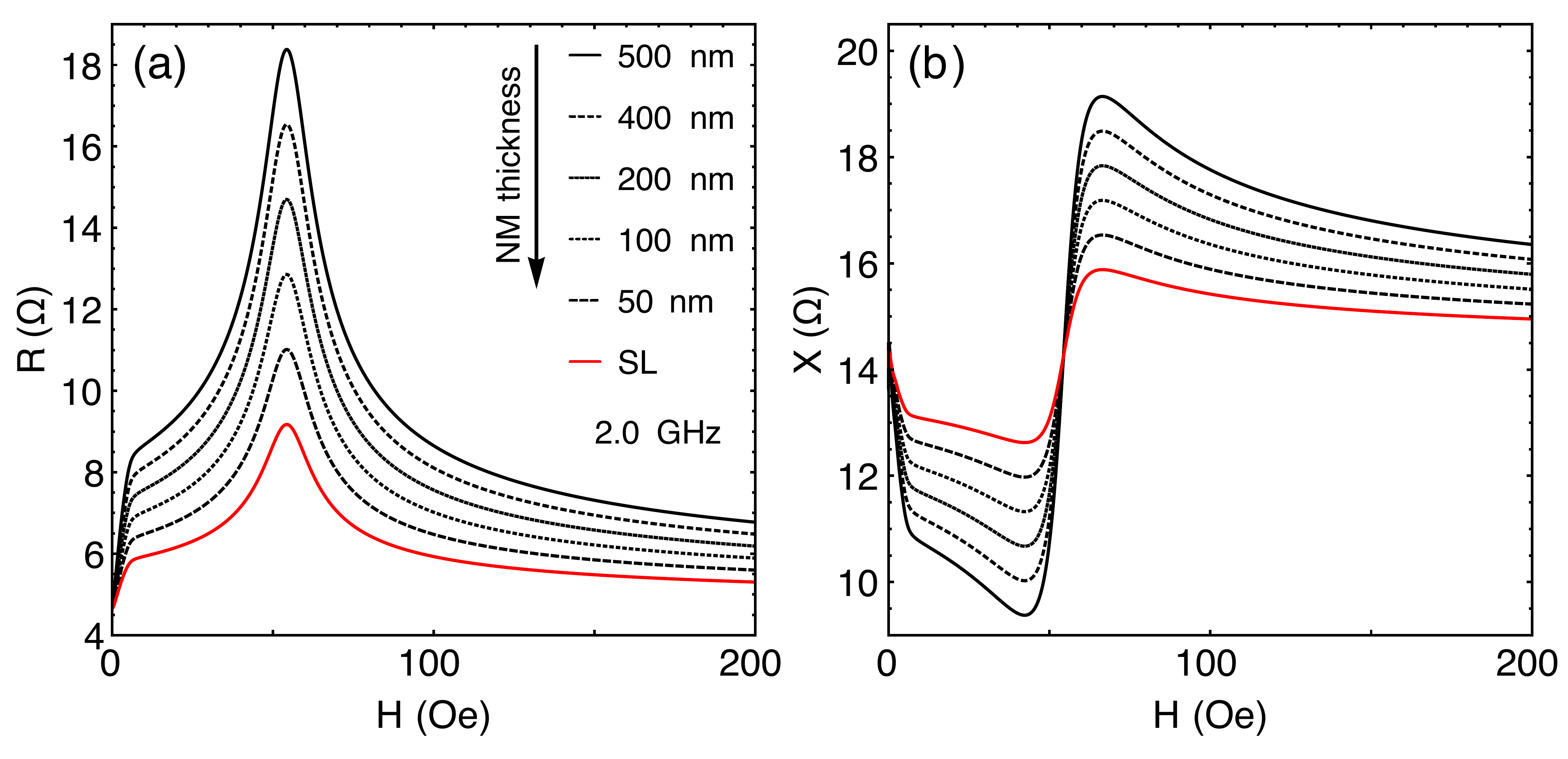}
\caption {(a) Real $R$ and (b) imaginary $X$ components of the longitudinal impedance, as a function of the external magnetic field at $2$ GHz, calculated for multilayered systems with different values of the thickness $t_1$ of the metallic non-magnetic layer. For $t_1= 0$, the multilayered system becomes a single layered system. The numerical calculations are performed using the same parameters previously employed for multilayered systems.}
\label{Fig_05}
\end{figure}

The primary difference between the results for single layered or multilayered systems resides basically in the amplitude of the MI curves. 
This fact is verified in Fig.~\ref{Fig_04} and evidenced in Fig.~\ref{Fig_05}. In this case, the MI variations are amplified for the multilayered system, a fact directly associated to the insertion of a metallic non-magnetic layer with high electric conductivity $\sigma_1$~\cite{JPDAP43p295004} and thickness $t_1$.

Regarding the electric properties and the conductivity of the system, it is well-known that multilayered systems present a clear dependence of the MI variations with the $\sigma_1/\sigma_2$ ratio. This behavior has been verified and detailed discussed in Ref.~\cite{SAAP81p71}, as well as also previously calculated by our group for a trilayered system~\cite{JAP110p093914}. 

Concerning the size of the system, the MI variations are strongly dependent on the thickness $t_1$ of the metallic non-magnetic layer, as shown in Fig.~\ref{Fig_05}. The higher MI variation values are verified for the thicker systems, with large $t_1$ values. This fact is due to the reduction of the electric resistance of the whole system with the increase of $t_1$, which is affected for both the increase of the system cross section and higher conductivity of the system. On the other hand, theoretically, in the limit of $t_1 \rightarrow 0$, Eq.~(\ref{impmult}) for the impedance is reduced to Eq.~(\ref{sl}), as expected~\cite{JMMM320pe25}, since the multilayered system becomes a single layered system for $t_1 = 0$.
%Bilinear and biquadratic interactions are not taken into account because we consider $t_1 = 0$ and $t_1 > 50$ nm, thickness range where they are not verified.

In this line, these numerical calculation results obtained for multilayered systems with different $\sigma_1/\sigma_2$ ratio or $t_1$ values are in qualitative concordance with several experimental results found in literature for multilayered films~\cite{ SAA129p256, APL94p042501, APL96p232501, JPDAP43p295004, TSF520p2173, JMMM355p136, JAP115p103908, APL104p102405, APL105p102409}.

The damping parameter $\alpha$ is also an important element for the determination of magnetoimpedance because of its relationship with the magnetization dynamics at high frequencies. Experimentally, the $\alpha$ value is influenced by the kind of the employed ferromagnetic material~\cite{APEX6p093001}, structural character~\cite{APEX6p093001}, and structure of the sample (single layered, multilayered, sandwiched samples). From the numerical calculations, we carry out an analysis similar to that presented by Kraus~\cite{JMMM195p764}, although here we consider a higher frequency range, where FMR signatures can be verified in the MI results. 

\begin{figure}[!]
\includegraphics[width=8.5cm]{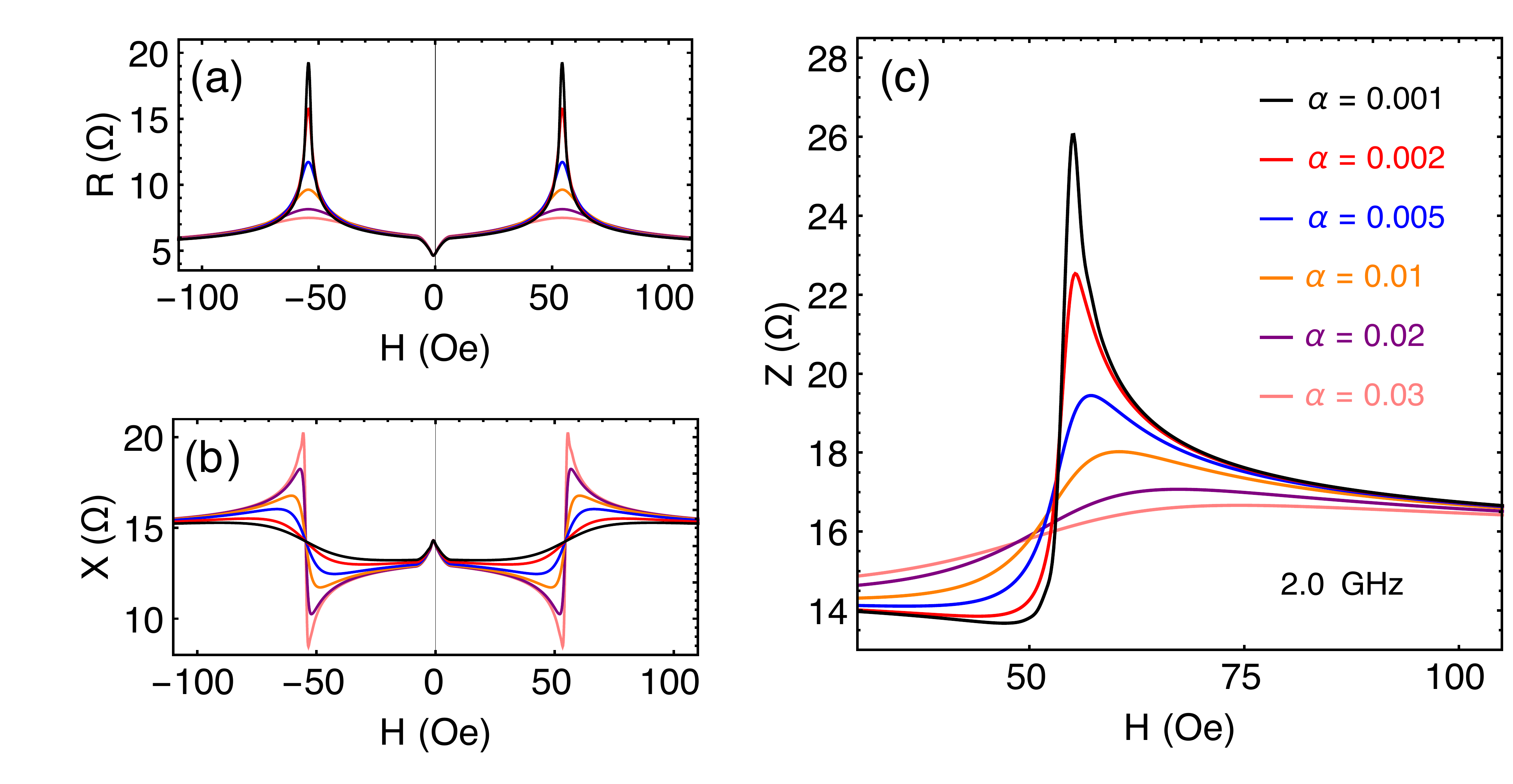}
\caption {(a) Real $R$ component, (b) imaginary $X$ component, and (c) impedance $Z$, as a function of the external magnetic field at $2$ GHz, calculated for multilayered systems with different values of the damping parameter $\alpha$. The numerical calculations are performed using the same parameters previously employed for multilayered systems.}
\label{Fig_06}
\end{figure}

Figure~\ref{Fig_06} presents the numerical calcultations of the real $R$ and imaginary $X$ components of the longitudinal impedance, as well as the impedance $Z$, for multilayered systems with different values of the damping parameter $\alpha$. Here, it can be clearly noticed that the amplitude of $R$, $X$, and $Z$ increases as the $\alpha$ value decreases. Moreover, a displacement of the peak position in field is observed when different $\alpha$ values are considered. This displacement leads to changes in the FMR frequency for a given external magnetic field and, therefore, it modifies the frequency limit between the regimes where distinct mechanisms are responsible for the MI effect variations. The features verified in these numerical calculations are present in experimental results measured in films with low damping parameter $\alpha$ values~\cite{APEX6p093001}.

\subsection{Exchange biased system}
\label{Numerical_calculation_exchange_bias}

Finally, we perform the numerical calculation for the longitudinal MI effect for an exchange biased system, as presented in Fig.~\ref{Fig_01}(d). 

The exchange biased system is composed by a ferromagnetic layer directly coupled to an antiferromagnetic layer. The sample configuration favors the appearance of the exchange interaction in the ferromagnetic/antiferromagnetic interface, described through a bias field $\vec H_{EB}$~\cite{APL94p042501, JMMM192p203}. Thus, the magnetic free energy density can be write as~\cite{APL94p042501}
\begin{equation}
\label{eb}
\xi = - \vec{M} \cdot \vec H - \frac{H_{k}}{2 M_s} \left(  \vec{M} \cdot \hat{u}_{k} \right)^2+4 \pi M^{2}_{s} \left( \hat{M} \cdot \hat{n} \right)- \vec{M} \cdot \vec H_{EB}.
\end{equation}  

For the numerical calculation for an exchange biased system, we consider the following parameters previously employed: $M_s = 780$ emu/cm$^3$, $H_k = 5$ Oe, $\theta_{k} = 90^{\circ}$, variable $\varphi_{k}$, $\alpha = 0.018$, $\gamma_G / 2 \pi = 2.9$ MHz/Oe, thickness of the ferromagnetic layer $t=500$ nm, $\theta_{H} = 90^{\circ}$, and $\varphi_{H}=90^\circ$. Beyond the traditional parameters, $H_{EB} = 50$ Oe, oriented along $\hat u_k$. In particular, the thickness of the antiferromagnetic layer is not considered for the numerical calculations.

\begin{figure}[!]
\includegraphics[width=8.5cm]{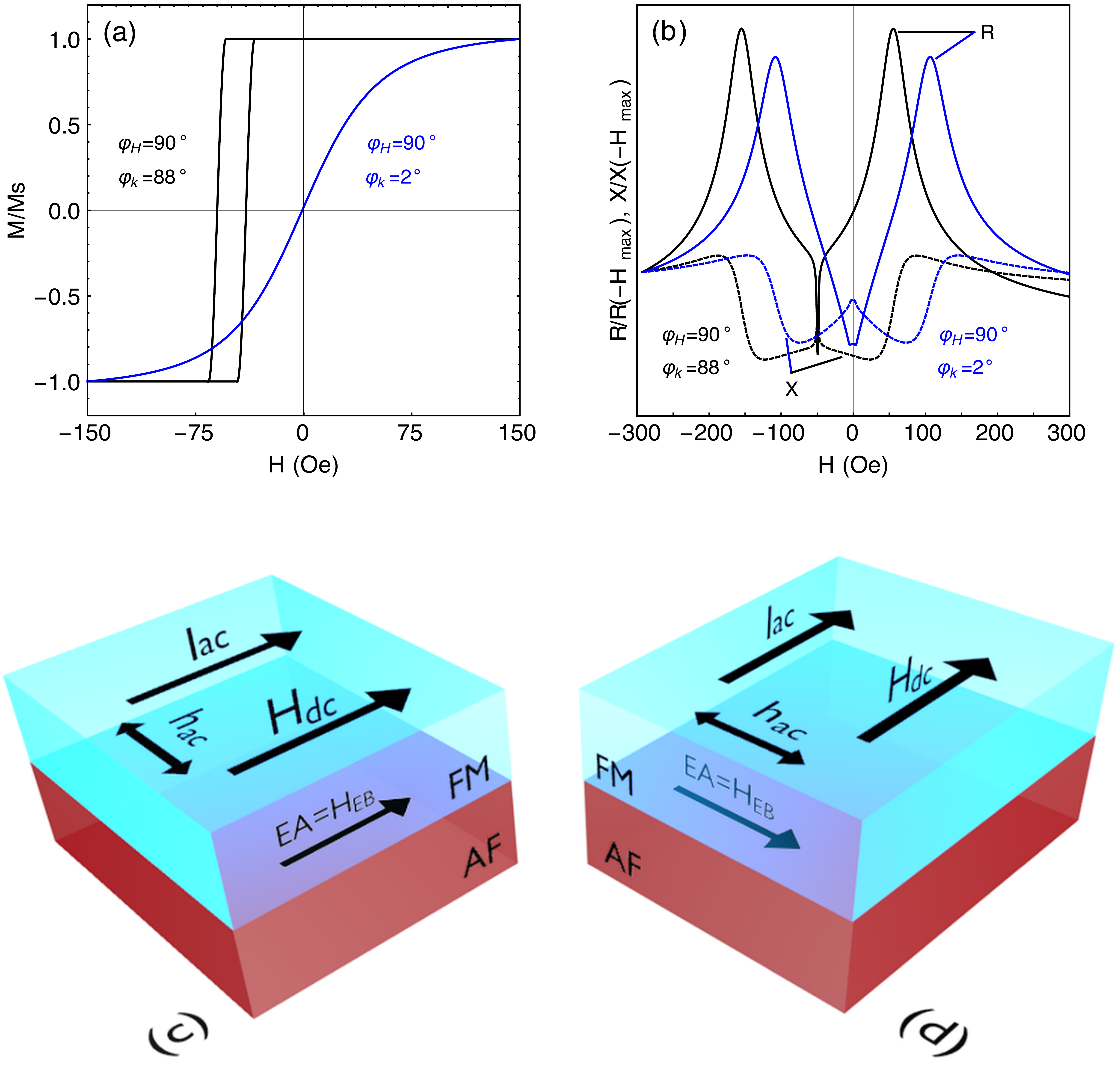}
\caption {(a) Normalized magnetization curves and (b) real $R$ and imaginary $X$ components of the longitudinal impedance $Z$ as a function of the external field at $2$ GHz, calculated for a exchange biased system when EA and $\vec H_{EB}$ are parallel ($\varphi_k = 88^\circ$) and perpendicular ($\varphi_k = 2^\circ$) to $\vec H$ and $I_{ac}$. Notice that to perform the numerical calculations for distinct orientations, $\varphi_k$ is modified, since we define $\mu_{xx}$ as the transverse magnetic permeability. The numerical calculations are performed for an exchange biased system system with $M_s = 780$ emu/cm$^3$, $H_k = 5$ Oe, $\theta_{k} = 90^{\circ}$, variable $\varphi_{k}$, $H_{EB} = 50$ Oe with $\vec H_{EB}$ oriented along $\hat u_k$, $\alpha = 0.018$, $\gamma_G / 2 \pi = 2.9$ MHz/Oe, $t=500$ nm, $\theta_{H} = 90^{\circ}$, and $\varphi_{H}=90^\circ$. Schematic representation of an exchange biased system and two configurations of the external and alternate magnetic fields and current sense when the easy magnetization axis (EA) and $\vec H_{EB}$ are (c) parallel and (d) perpendicular to $\vec H$ and $I_{ac}$.}
\label{Fig_07}
\end{figure}

Figure~\ref{Fig_07} shows the numerical calculations for the normalized magnetization curves and real $R$ and imaginary $X$ components obtained as a function of the external magnetic field at $2$ GHz for two different orientations between $\vec H_{EB}$ and $\hat u_k$ with $\vec H$ and $I_{ac}$, together with the schematic representations of the two configurations. In particular, in this case the calculations are performed considering the Eqs.~(\ref{eb}), (\ref{muxx}), and (\ref{sl}).

Considering the magnetization curves (See Fig.~\ref{Fig_07}(a)), the exchange bias can be clearly identified through the shift of the curve, where the maximum exchange bias field is observed when $\vec H \parallel \vec H_{EB}$ (Fig.~\ref{Fig_07}(c)), as expected. As the angle between $\vec H_{EB}$ and $\hat u_k$ with $\vec H$ and $I_{ac}$ is increased, a reduction of the component of the exchange bias field along $\vec H$ is verified, evidenced by the decrease of the shift (Not shown here). For the limit case of $\vec H \perp \vec H_{EB}$ (Fig.~\ref{Fig_07}(d)), none shift of the curve is observed. At the same time, an evolution of the shape of the magnetization curve is noticed as the angle increases. These exchange bias features are reflected in the behavior of the MI curves. In particular, the shift of the curves of the real and imaginary components of the impedance follows the one of the respective magnetization curve (See Fig.~\ref{Fig_07}(b)).

\begin{figure}[!]
\includegraphics[width=8.5cm]{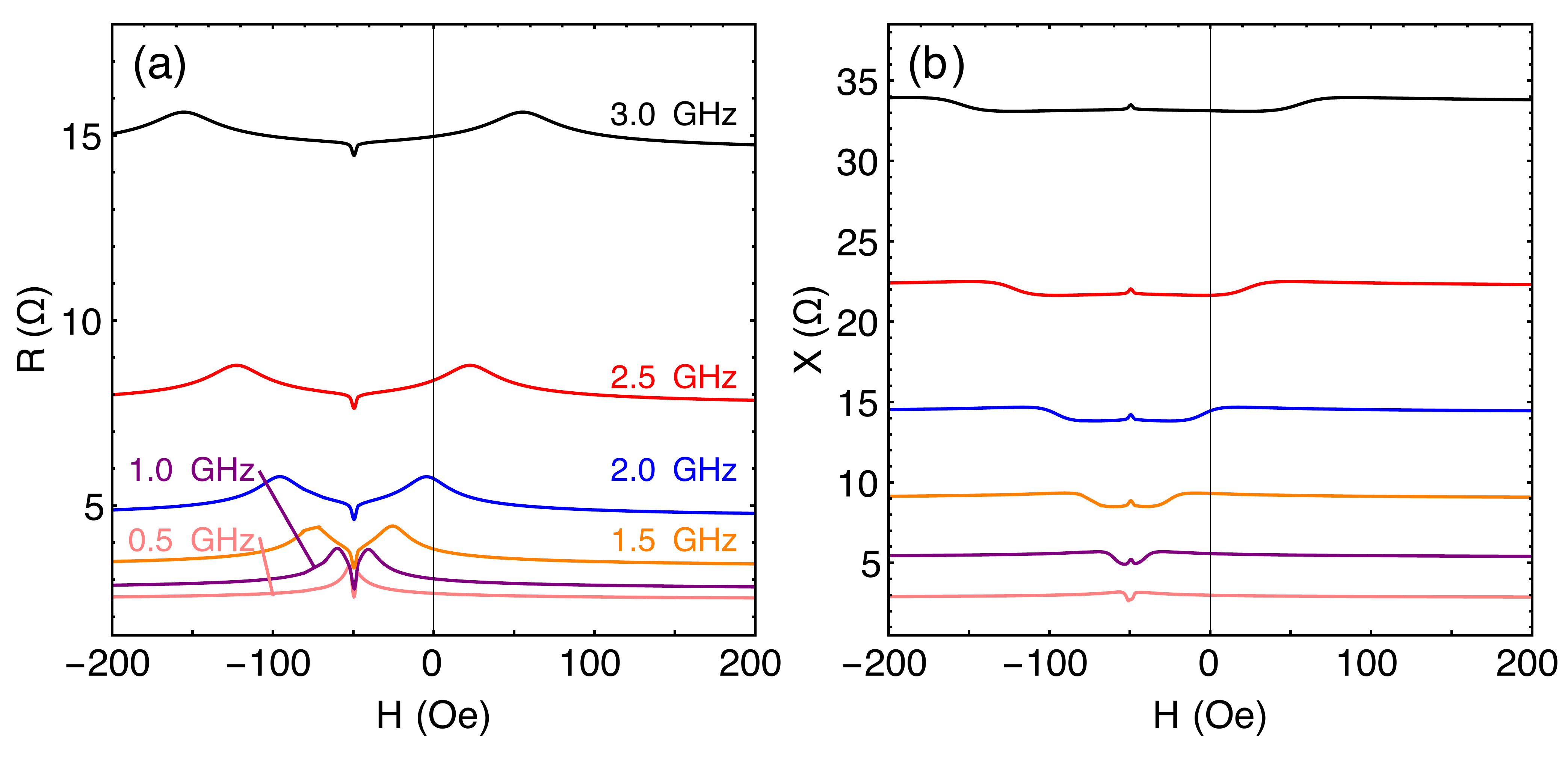}
\caption {(a) Real $R$ and (b) imaginary $X$ components of the longitudinal impedance as a function of the external magnetic field for selected frequency values calculated for an exchange biased system with the same parameter employed previously and $\varphi_k = 88^\circ$.}
\label{Fig_08}
\end{figure}

Figure~\ref{Fig_08} shows the numerical calculations for the real $R$ and imaginary $X$ components of the longitudinal impedance as a function os the external magnetic field for selected frequency values, calculated for an exchange biased system for the configuration of $\vec H_{EB} \parallel \vec H$.
For exchange biased systems, the well-known symmetric magnetoimpedance behavior around $H = 0$ for anisotropic systems~\cite{APL67p857} is entirely shifted to $H =H_{EB}$~\cite{APL94p042501}. Besides, the MI curves reflect all classical features of the magnetoimpedance in systems without the exchange bias, including the $Z$ behavior for distinct orientation between the anisotropy and external magnetic field~\cite{SAAP106p187}, as well as the $R$, $X$ and $Z$ behavior with frequency~\cite{JMMM195p764,APL69p3084,JAP110p093914}, together with the new features owed to exchange bias effect~\cite{APL94p042501}. In this case, a single peak placed at $H = H_{EB}\pm H_{c}$, where $H_c$ is the coercive field, can be observed from $0.15$ GHz (not presented here) up to $0.5$ GHz, and it is due to changes in the transverse magnetic permeability. The single peak becomes more pronounced with the increase of the frequency. At around $0.6$ GHz, the single peak splits in a double peak structure symmetric at $H=H_{EB}$. In classical MI experiments, this evolution of the curves from a single peak to a double peak structure is verified when both the external magnetic field and electrical current are applied along the easy magnetization axis~\cite{APL67p857}, and is owed to the typical shape of FMR dispersion relation for this geometry~\cite{JMMM195p764, SAAP106p187}. 
%On the other hand, for $\vec H \perp \vec H_{EB}$, a double peak behavior, symmetrical around $H= 0$ ($H_{EB}\sim 0$, and $H_c \sim 0$), is present for the whole frequency range, again a feature of FMR relation dispersion~\cite{JMMM195p764, SAAP106p187, JAP110p093914}, in a signature of the parallel alignment of the external magnetic field and {\it ac} current along of the hard magnetization axis. At the same time, the displacement of the peak position in the double peak structure toward higher fields as the frequency is increased is verified, following the behavior predicted for the FMR effect~\cite{APL69p3084, JMMM195p764, SAAP106p187, JAP110p093914}.

These numerical calculation results obtained for exchange biased systems are in qualitative agreement with experimental results found in literature for ferromagnetic films with exchange bias~\cite{APL94p042501, APL96p232501, APL104p102405}.

\section{Comparision with the experiment}

The previous tests performed with the theoretical approach have qualitatively described the main features of {\it single layered, multilayered and exchange biased systems}. To verify the validity of the theoretical approach, we investigate the quasi-static and dynamical magnetic properties of an exchange biased multilayered film and compare the experimental results with numerical calculations obtained a  {\it exchange biased multilayered system}. The complexity of the considered system, including different features previously studied, and the quantitative agreement with experimental results do confirm the robustness of our theoretical approach.

\subsection{Experiment}

Here, we investigate a [Ni$_{20}$Fe$_{80}$ ($40$ nm)/Ir$_{20}$Mn$_{80}$($20$ nm)/Ta($1$ nm)]$\times 20$ ferromagnetic exchange biased multilayered film. The film is deposited by magnetron sputtering onto a glass substrate, covered with a $2$ nm-thick Ta buffer layer. The deposition process is performed with the following parameters: base vacuum of $8.0 \times 10^{-8}$ Torr, deposition pressure of $5.0$ mTorr with a $99.99$\% pure Ar at $50$ sccm constant flow, and DC source with current of $150$ mA for the deposition of the Ta and IrMn layers, as well as $65$ W set in the RF power supply for the deposition of the NiFe layers. With these conditions, the obtained deposition rates are $0.08$ nm/s, $0.67$ nm/s and $0.23$ nm/s for NiFe, IrMn and Ta, respectively. During the deposition, the substrate with dimensions of $5 \times 2$ mm$^2$ is submitted to a constant magnetic field of $2$ kOe, applied along the main axis of the substrate in order to define an easy magnetization axis and induce a magnetic anisotropy and an exchange bias field $\vec H_{EB}$ in the interface between the NiFe and IrMn layers.

Quasi-static magnetization curves are obtained with a vibrating sample magnetometer, measured along and perpendicular to the main axis of the films, in order to verify the magnetic behavior.

The magnetoimpedance effect is measured using a RF-impedance analyzer Agilent model $E4991$, with $E4991A$ test head connected to a microstrip in which the sample is the central conductor, which is separated from the ground plane by the substrate. The electric contacts between the sample and the sample holder are made with $24$ h cured low resistance silver paint. To avoid propagative effects and acquire just the sample contribution to MI, the RF impedance analyzer is calibrated at the end of the connection cable by performing open, short, and load ($50$ $\Omega$) measurements using reference standards. The probe current is fed directly to one side of the sample, while the other side is in short circuit with the ground plane. The {\it ac} current and external magnetic field are applied along the length of the sample. MI measurement is taken over a wide frequency range, between $0.5$ GHz and $3.0$ GHz, with maximum applied magnetic fields of $\pm 350$ Oe. While the external magnetic field is swept, a $0$ dBm ($1$ mW) constant power is applied to the sample characterizing a linear regime of driving signal. Thus, at a given field value, the frequency sweep is made and the real $R$ and imaginary $X$ parts of the impedance are simultaneously acquired. For further information on the whole procedure, we suggest Refs.~\cite{JPDAP41p175003, JAP101p033908}. The curves are known to exhibit hysteretic behavior, associated with the coercive field. However, in order to clarify the general behavior, only curves depicting the field going from negative to positive values are presented.

\subsection{Results}

\begin{figure}[!]
\includegraphics[width=8.5cm]{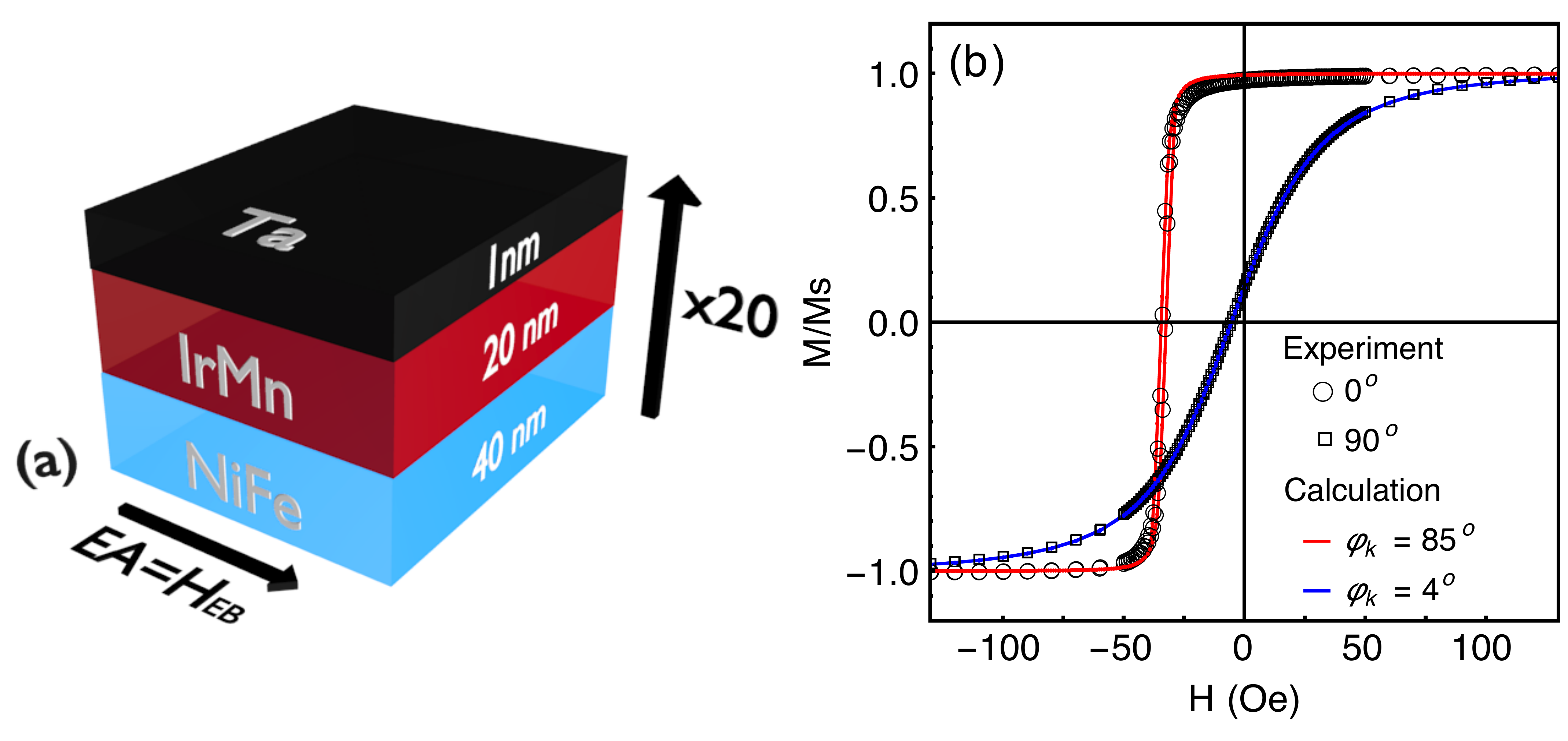}
\caption {(a) Schematic diagram of an exchange biased multilayered system. Experimentally, we produce a [Ni$_{20}$Fe$_{80}$ ($40$ nm)/Ir$_{20}$Mn$_{80}$($20$ nm)/Ta($1$ nm)]$\times 20$ ferromagnetic multilayered film, in which the the easy magnetization axis EA and the exchange bias field $\vec H_{EB}$ are oriented in the same direction. (b) Normalized magnetization curves obtained experimentally when $\vec H$ is applied along ($0^\circ$) and perpendicular ($90^\circ$) to the main axis of the film, together with numerical calculations performed for an exchange biased multilayered system with $\varphi_{ki} = 85^\circ$ and $\varphi_{ki} = 4^\circ$, respectively. Notice that, for all calculations, with $\theta_{H} = 90^{\circ}$, and $\varphi_{H}=90^\circ$. The other parameters of the system employed in the numerical calculation are $M_{si} = 780$ emu/cm$^3$, $H_{ki} = 2$ Oe, $\theta_{ki} = 90^{\circ}$, variable $\varphi_{ki}$, $H_{EB} = 30.5$ Oe with $\vec H_{EB}$ oriented along $\hat u_{ki}$, $\alpha = 0.018$, $\gamma_G / 2 \pi = 2.73$ MHz/Oe~\cite{PRB38p6847}, $t_1= 1$ nm, $t_2 = 40$ nm, $\sigma_1 = 6 \times 10^7\,(\Omega\textrm{m})^{-1}$, and $\sigma_2 = \sigma_1/0.5$.}
\label{Fig_09}
\end{figure}

We perform numerical calculation for the quasi-static and dynamical magnetic properties of an exchange biased multilayered system, as shown in Fig.~\ref{Fig_09}(a).

To model the exchange biased multilayered system, we consider a magnetic free energy density that can be written as
\begin{equation}
\label{MEB} \small
 \xi = \sum_{i=1}^{20} \left[
 \begin{aligned}
&- \vec{M}_i \cdot \vec{H} - \frac{H_{ki}}{2 M_{si}} \left(  \vec{M}_{i}\cdot \hat{u}_{ki} \right)^2 \\
& +4 \pi M^{2}_{si} \left( \hat{M}_{i} \cdot \hat{n} \right) - \vec{H}_{EB} \cdot \vec{M_i}
\end{aligned}
\right].
\end{equation}

With respect to numerical calculations, the following parameters must be defined to describe the experimental magnetization and MI curves: magnetization and saturation magnetization of each ferromagnetic layer, $M_i$ and $M_{si}$ respectively, uniaxial anisotropy field $H_{ki}$, uniaxial anisotropy versor $\hat u_{ki}$, exchange bias field $H_{EB}$, thicknesses, $t_1$ and $t_2$, and conductivities, $\sigma_1$ and $\sigma_2$, of the of the non-magnetic and ferromagnetic layers, respectively, damping parameter $\alpha$, gyromagnetic factor $\gamma_G$, and external magnetic field $\vec H$. The thickness of the antiferromagnetic layer is not considered for the numerical calculations, however, experimentally, it is thick enough to neglect the bilinear and biquadratic coupling between the ferromagnetic layers.

The calculation of the magnetization curves is carried out using the same minimization process developed for the MI calculation, without the $\vec h_{ac}$ field. This process consists in to determine the $\theta_{M}$ and $\varphi_{M}$ values that minimize the magnetic free energy density for the studied system for each external magnetic field value. Thus, since the calculated magnetization curve validates the experimental magnetization behavior, the aforementioned parameters are fixed to perform the numerical calculations of MI behavior. As previously cited, there is an offset increase in the real and imaginary parts of the experimental impedance as a function of frequency, a feature of the electrical/metallic contribution to MI that is not taken into account in theoretical models. Thus, it is inserted in the MI numerical calculation from the fitting of the measured $R$ and $X$ curves as a function of the frequency for the highest magnetic field value~\cite{JAP110p093914}.

Figure~\ref{Fig_09}(b) shows the normalized magnetization curves of the produced exchange biased multilayered multilayered film. Experimental magnetization curves are obtained along two different directions, when $\vec H$ is applied along and perpendicular to the main axis of the films. It is important to point out that a constant magnetic field is applied along the main axis during the deposition process. As a matter of fact, by comparing experimental curves, it is possible to observe that magnetic anisotropy is induced during the film growth, confirming an easy magnetization axis and an exchange bias field oriented along the main axis of the film. 

From the magnetization curve measured along the main axis of the film, we find the coercive field $\sim 2$ Oe and $H_{EB}\sim 30$ Oe. Thus, to the numerical calculation, we consider the following parameters $M_{si} = 780$ emu/cm$^3$, $H_{ki} = 2$ Oe, $\theta_{ki} = 90^{\circ}$, variable $\varphi_{ki}$, $H_{EB} = 30.5$ Oe with $\vec H_{EB}$ oriented along $\hat u_{ki}$, $\alpha = 0.018$, $\gamma_G / 2 \pi = 2.73$ MHz/Oe~\cite{PRB38p6847}, $t_1= 1$ nm, $t_2 = 40$ nm, $\sigma_1 = 6 \times 10^7\,(\Omega\textrm{m})^{-1}$, and $\sigma_2 = \sigma_1/0.5$. Since $\theta_{H} = 90^{\circ}$ and $\varphi_{H}=90^\circ$, we obtain $\varphi_{ki} = 85^\circ$ and $\varphi_{ki} = 4^\circ$, respectively, for the two measurement directions. A small misalignment between anisotropy and field can be associated to stress stored in the film as the sample thickness increases~\cite{JAP101p033908}, as well as small deviations in the sample position in an experiment are reasonable. This is confirmed through the numerical calculation of the magnetization curves. Notice the striking quantitave agreement between experiment and theory.

\begin{figure*}[!]
\includegraphics[width=.8\textwidth]{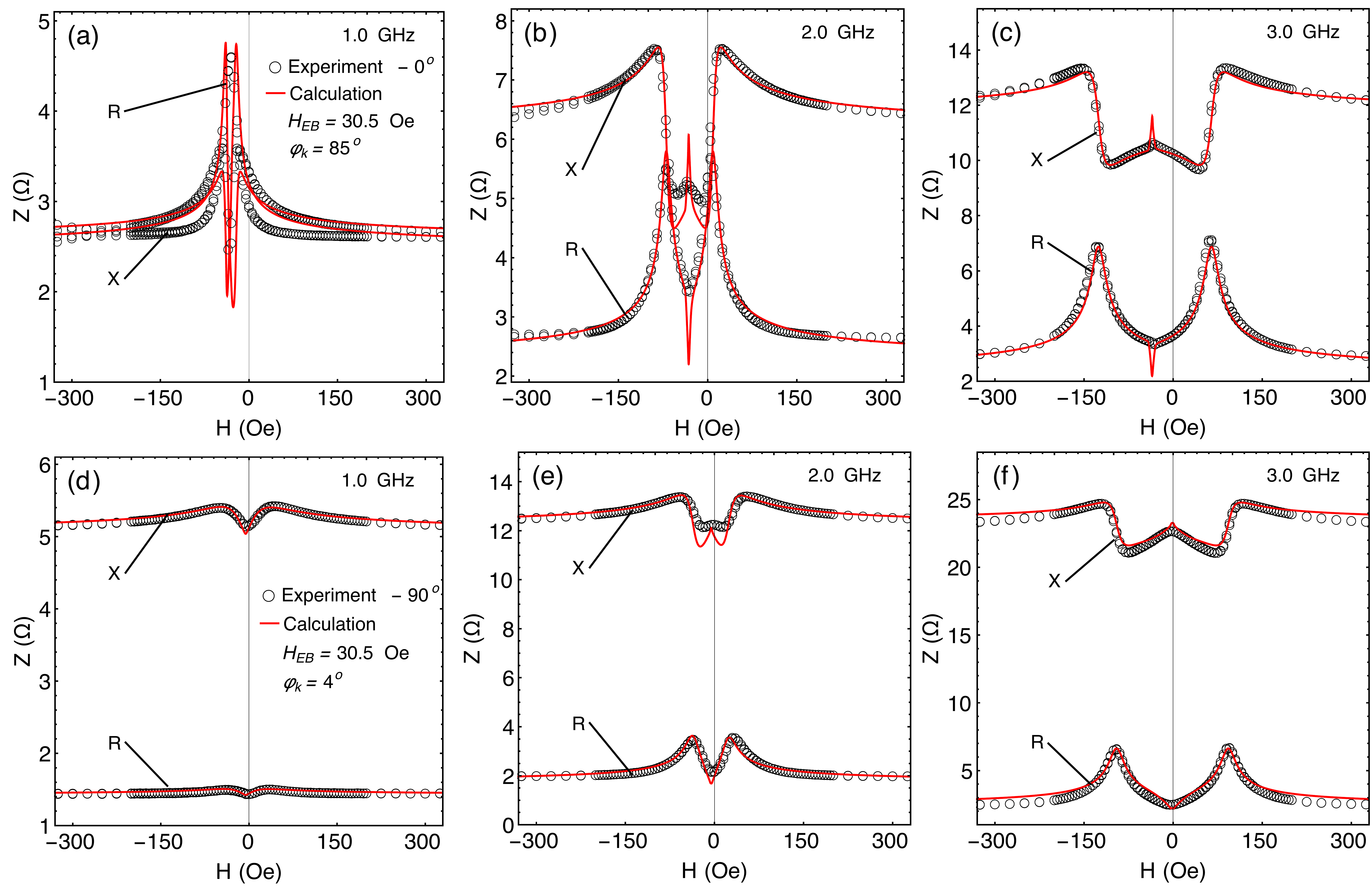}
\caption {Experimental results and numerical calculation of real $R$ and imaginary $X$ components of the longitudinal impedance as a function of the external magnetic field for selected frequency values. (a)-(c) Experimental data obtained when $\vec H$ is applied along ($0^\circ$) to the main axis of the film, together with numerical calculations performed for an exchange biased multilayered system with $\varphi_{ki} = 85^\circ$. (d)-(f) Similar plot of experimental data when the field is perpendicular ($90^\circ$) to the main axis of the film, with numerical calculations performed with $\varphi_{ki} = 4^\circ$. The parameters employed in the numerical calcutation are the ones fixed from the calculation of the magnetization curves. In this case, they are: $M_{si} = 780$ emu/cm$^3$, $H_{ki} = 2$ Oe, $\theta_{ki} = 90^{\circ}$, variable $\varphi_{ki}$, $H_{EB} = 30.5$ Oe with $\vec H_{EB}$ oriented along $\hat u_{ki}$, $\alpha = 0.018$, $\gamma_G / 2 \pi = 2.73$ MHz/Oe~\cite{PRB38p6847}, $t_1= 1$ nm, $t_2 = 40$ nm, $\sigma_1 = 6 \times 10^7\,(\Omega\textrm{m})^{-1}$, $\sigma_2 = \sigma_1/0.5$, $\theta_{H} = 90^{\circ}$ and $\varphi_{H}=90^\circ$.}
\label{Fig_10}
\end{figure*}

As mentioned above, parameters are fixed from the calculation of magnetization curves and used to describe the MI behavior. Thus, from Eqs.~(\ref{eml}), (\ref{muxx}) and (\ref{impmult}), the real $R$ and imaginary $X$ components of the longitudinal impedance as a function of field and frequency for an exchange biased multilayered system can be calculated. Fig.~\ref{Fig_10} shows experimental data and numerical calculation of $R$ and $X$ as a function of $H$ for selected frequency values for both considered directions.

For all cases, it is evident the quantitative agreement between the experimental data and numerical calculation. In particular, the numerical calculations performed using the considered magnetic permeability and MI models, with parameters fixed by analyzing the magnetization curves, are able to describe all the main features of each impedance component for the whole frequency range.

Although it is well-known that the changes of magnetic permeability and impedance with magnetic fields at different frequency ranges are caused by three distinct mechanisms~\cite{PMS53p323,IEEETM29p1245,APL64p3652,APL69p3084}, the determination of the precise frequency limits between regimes is a hard task, since the overlap of contributions to MI of distinct mechanisms, such as the skin and FMR effects, is very likely to occur. Thus, the use of distinct models for magnetic permeability and their use in calculating MI become restricted, since it is not possible to determine when to leave one model and start using another one as the frequency is changing.

Even there are distinct mechanisms controlling MI variations at different frequency ranges, all of our experimental findings are well described by the theoretical results calculated using the aforementioned magnetic permeability and MI models. This is due to the fact that the distinct mechanism contributions at different frequency ranges are included naturally in the numerical calculation through magnetic permeability.

%The microstrip contribution was obtained adjusting the real and imaginary components of the impedance in frequency function for a high external magnetic field value (here the sample is saturated). Thereby, the electrical contribution was added in the MI theoretical approach. Despite the measured frequency range was from $0.5$ GHz up to $3.0$ GHz the Fig. \ref{figure11} shows the experimental end theoretical model for $1.0$, $2.0$ and $3.0$ GHz. 

%In particular for ML$_{90}$ sample, the discrepancy at $H_{EB} \approx 30.5$ Oe can be associated with the Gilbert damping constant value choose for this calculation, as well as, with a misalignment between the cut and the induced anisotropy during the growth of the matrix sample. 

\section{Conclusion}
\label{Conclusion}

As an alternative to the traditional FMR experiment, the magnetoimpedance effect corresponds as a promissing tool to investigate ferromagnetic materials, revealing aspects on the fundamental physics associated to magnetization dynamics, broadband magnetic properties, important issues for current and emerging technological applications for magnetic sensors, as well as insights on ferromagnetic resonance effect at non-saturated magnetic states. In this sense, its study in ferromagnetic samples with distinct features becomes a very important task.

In this paper, we perform a theoretical and experimental investigation of the magnetoimpedance effect for the thin film geometry in a wide frequency range. 

In particular, we calculate the longitudinal magnetoimpedance for single layered, multilayered or exchange biased systems from an approach that considers a magnetic permeability model for planar geometry and the appropriate magnetic free energy density for each structure. Usually, theoretical models that describe magnetization dynamical properties and the MI of a given system consider more than one approach to magnetic permeability. This is due to the fact that these permeability approaches reflect distinct mechanisms responsible for MI changes, applicable only for a restricted range of frequencies where the mechanism is observed. Thus, the difficult task of choosing the correct magnetic permeability model to use at a certain frequency range explains the reduced number of reports comparing MI experimental results and theoretical predictions for a wide frequency range. Anyway, even there are distinct mechanisms controlling MI variations at different frequency ranges, with the magnetic permeability and MI models considered here, the distinct mechanism contributions at different frequency ranges are included naturally in the numerical calculation through magnetic permeability. For this reason, the numerical calculations for different systems succeed to describe the main features of the MI effect in each structure, in concordance with experimental results found in literature.

At the same time, we perform experimental magnetization and MI measurements in a multilayered film with exchange bias. To interpret them, numerical calculations are performed using the described magnetic permeability and MI models. With parameters fixed by analyzing the magnetization curves, quantitative agreement between the experimental MI data and numerical calculation is verified, and we are able to describe all the main features of each impedance component for the whole frequency range. Thus, we provide experimental evidence to confirm the validity of the theoretical approach to describe the magnetoimpedance in ferromagnetic films.

Although we perform here all the analysis just for an exchange biased multilayered film, since a general model is used to describe magnetic permeability, it can be considered in the study of samples with any planar geometry, such as films, ribbons and sheets, given that an appropriate magnetic free energy density and adequate MI model are considered. In this sense, the simplicity and robustness place this theoretical approach as a powerful tool to investigate the permeability and longitudinal magnetoimpedance for the thin film geometry in a wide frequency range. 

In particular, we focus on the $\mu_{xx}$ term of the magnetic permeability tensor and on the longitinal magnetoimpedance. This is due to the fact that our experimental setup provides information related to the transverse magnetic permeability. On the other hand, the $\mu_{yy}$, $\mu_{zz}$, and off-diagonal terms of the magnetic permeability tensor can bring relevant information on the MI effect, since the correct impedance expression is obtained. At the same time, this information can be measured by considering a distinct experimental system. These next steps are currently in progress.

\begin{acknowledgments} 
The research is partially supported by the Brazilian agencies CNPq (Grants No.~$310761$/$2011$-$5$, No.~$476429$/$2010$-$2$, and No.~$555620$/$2010$-$7$), CAPES, FAPERJ, and FAPERN (Grant PPP No.~$013$/$2009$, and Pronem No.~$03/2012$). M.A.C. and F.B. acknowledge financial support of the INCT of Space Studies.
\end{acknowledgments}

\end{document}